\documentclass[a4paper,fleqn,usenatbib]{mnras}

\usepackage[T1]{fontenc}
\usepackage{ae,aecompl}

\usepackage{graphicx}	% Including figure files
\usepackage{amsmath}	% Advanced maths commands
\usepackage{amssymb}	% Extra maths symbols

\usepackage{hyperref}   % Make the eprint and doi bibtex fields hyperlinks
\usepackage{caption}    % Allows line break inside caption
\newcommand{\eff}{_{\textrm{\tiny eff}}} % non-italic subscript "eff"
\newcommand{\subsun}{_{\sun}} % subscript solar mass symbol
\title[The sdA problem I: Physical Properties]{The sdA problem I: Physical Properties}

\author[I. Pelisoli et al.]{
Ingrid Pelisoli$^{1,2}$\thanks{E-mail: ingrid.pelisoli@ufrgs.br},
S. O. Kepler$^{1}$,
D. Koester$^{3}$,
\\
$^{1}$Instituto de F\'{i}sica, Universidade Federal do Rio Grande do Sul, 91501-900 Porto-Alegre, RS, Brazil\\
$^{2}$Department of Physics, University of Warwick, Coventry CV4 7AL, UK\\
$^{3}$Institut f\"{u}r Theoretische Physik und Astrophysik, Universit\"{a}t Kiel, D-24098 Kiel, Germany\\
}

\date{Accepted XXX. Received YYY; in original form ZZZ}

\pubyear{2017}

\begin{document}
\label{firstpage}
\pagerange{\pageref{firstpage}--\pageref{lastpage}}
\maketitle

\begin{abstract}
The so-called sdA stars are defined by having H-rich spectra and surface gravities similar to hot subdwarf stars, but effective temperature below the zero-age horizontal branch. Their evolutionary history is an enigma: their surface gravity is too high for main sequence stars, but too low for single evolution white dwarfs. They are most likely byproducts of binary evolution, including blue-stragglers, extremely-low mass white dwarf stars (ELMs) and their precursors (pre-ELMs). A small number of ELMs with similar properties to sdAs is known. Other possibilities include metal-poor A/F dwarfs, second generation stars, or even stars accreted from dwarf galaxies. In this work, we analyse colours, proper motions and spacial velocities of a sample of sdAs from the Sloan Digital Sky Survey to assess their nature and evolutionary origin. We define a probability of belonging to the main sequence and a probability of being a (pre-)ELM based on these properties. We find that 7 per cent of the sdAs are more likely to be (pre-)ELMs than main sequence stars. However, the spacial velocity distribution suggests that over 35 per cent of them cannot be explained as single metal-poor A/F stars.
\end{abstract}


%%%%%%%%%%%%%%%%%%%%%%%%%%%%%%%%%%%%%%%%%%%%%%%%%%

%%%%%%%%%%%%%%%%% BODY OF PAPER %%%%%%%%%%%%%%%%%%

\section{Introduction}

The evolution of single stars is a fairly well understood process. There are of course many uncertainties concerning specific phases \citep[e.g. the asymptotic giant branch,][]{miller2016} and processes \citep[e.g. convection,][]{bressan2013,tremblay2013} throughout the evolution, but relating an initial mass and metallicity to a final outcome can be done with reasonable precision. Only a very small amount of objects, much less than 5 per cent of all stars in the Galaxy, will end their lives as neutron stars or black holes --- those with initial masses larger than about 7--10.6~$M\subsun${} \citep[e.g.][]{woosley2015}. The remaining over 95 per cent stars will become white dwarf stars, whose evolution can be approximated by a slow cooling process. Hence, as they are not only abundant but also long-lived, white dwarfs are very useful in obtaining information of all galactic populations \citep{isern2001, liebert2005, bono2013, tremblay2016, kilic2017, cojocaru2017}.

White dwarf stars are also one possible result of binary evolution, both in binary systems \citep[e.g.][]{rmansergas2016} or as single stars resulting from merger events \citep{brown2016}. Of remarkable interest are the extremely-low mass white dwarfs \citep[ELMs, $M\lesssim0.20~M\subsun$, see e.g. the ELM Survey:][]{elmsurveyI, elmsurveyII, elmsurveyIII, elmsurveyIV, elmsurveyV, elmsurveyVI, elmsurveyVII}, which can only be formed in interacting binary systems within a Hubble time. The evolution of single stars can lead to white dwarfs with masses down to 0.30--0.45 solar masses \citep{kilic2007}, but main-sequence progenitors that would end up as lower mass white dwarfs have main sequence lifetimes exceeding the age of the Universe. Objects with $0.30 < M < 0.45~M\subsun$ are usually referred to as low-mass white dwarfs. Their binary fraction is still high, about 30 per cent \citep{brown2011}, because some form of enhanced mass-loss is needed to form them, and binarity is the easiest way to achieve this. Severe mass loss in the first ascent giant branch, attributed to high metallicity \citep{hansen2005}, can also lead to single low mass white dwarfs \citep{kilic2007}. The binary fraction below $0.2-0.3~M\subsun$, on the other hand, could be up to 100 per cent \citep{elmsurveyVII}, encompassing not only the ELMs but also the pre-ELMs, which still have not reached the white dwarf cooling track \citep{maxted2014}.

Less than a hundred ELMs are known to date, making it difficult to test and improve theoretical models \citep{corsico2014,corsico2016,istrate2016}. White dwarf catalogues such as \citet{dr7cat}, which relies on Sloan Digital Sky Survey (SDSS) data release (DR) 7, usually opt to remove any object with estimated surface gravity below 6.5, the single evolution limit, excluding the ELMs and pre-ELMs from their analysis. This flaw motivated \citet{dr12cat} to extend their catalogue down to $\log g=5.5$, which unveiled thousands of objects in the ELM range of $\log g$ and $T\eff \lesssim 20\,000$~K. As their nature as ELMs or pre-ELMs cannot be confirmed without probing their radius and verifying they are compact objects, they were dubbed {\it sdAs}, referring to their $\log g$ placing them below the main sequence as the subdwarfs, and their hydrogen-dominated A-type spectra. This, however, is merely a spectroscopic classification, saying nothing about the evolutionary nature of these objects, which remains a puzzle. They cannot be canonical He-core burning subdwarfs, which show $T\eff \geq 20\,000$~K, lying above the zero-age horizontal branch (ZAHB). Their estimated $\log g$ suggests they are not H core main sequence stars, seeing that evolutionary models indicate that the maximum $\log g$ of main sequence A stars is around 4.75 \citep[see][and references therein]{romero2015}.

\citet{hermes2017} studied the sdAs published by \citet{dr12cat}, which were modelled with pure-H atmosphere models, using radial velocities obtained from SDSS subspectra, photometric colours, and reduced proper motions, and concluded that over 99 per cent of them are unlikely to be ELMs. Likewise, \citet{brown2017} obtained follow-up time-resolved spectroscopy for five eclipsing systems and concluded they are not ELMs, but metal-poor $M\sim1.2~M\subsun$ main sequence stars with $M\sim0.8~M\subsun$ companions. They suggest that the majority of sdAs are metal-poor A--F type stars. Considering their distance modulus $(m-M)>14.5$ at the SDSS bright saturation, this puts them in the halo. Given a halo age of 11.5~Gyr \citep{kalirai2012, kilic2012, si2017, kilic2017}, only objects with $M\lesssim 0.8~M\subsun$ should still be in the main sequence \citep[e.g. as obtained with the {\sc lpcode} by][]{althaus2003}, and that assuming a very low metallicity of $Z=0.0001$, i.e. halo A-type stars should even have already evolved off the main sequence. For the same metallicity, main sequence lifetimes of A-stars are between 0.5--1.5~Gyr. Models by \citet{schneider2015} suggest that mass accretion can make a star appear up to 10 times younger than its parent population, what explains the so-called blue stragglers \citep[first identified by][]{sandage1953}. The sdAs could be explained as blue stragglers, when the $\log g$ is not higher than the main sequence limit.

\citet{brown2017} state that the $\log g$ derived from pure hydrogen models for sdA stars suffers from a systematic overestimate of $\sim 1$~dex on the surface gravities, likely explained by metal line blanketing below 9000~K. In this work we re-analyse the sdA sample selected in \citet{dr12cat} in the light of new spectral models including metals in solar abundances, first reported in \citet{pelisoli2017}. We assess the changes in $\log g$ between the two models, analysing particularly the sdAs in \citet{dr12cat}, to understand why \citet{hermes2017} and \citet{brown2017} found only a small percentage of them to have ELM properties. We extend the analysis to other sdAs selected from the SDSS database, identifying new pre-ELM and ELM candidates. Their colours, proper motions and spacial velocities are studied in order to assess their possible nature. The physical parameters we obtained are compared to both single evolution and binary evolution models to assess if they can be explained by this scenarios. Based on our findings, we estimate for each object in our sample a probability of being a (pre-)ELM and a probability of being a MS star. These probabilities can be used to guide future follow-up of these objects. More than one evolution channel will certainly be needed to explain the sdA population. The binaries within the sample can help us better understand binary stellar evolution and its possible outcomes, while the properties and dynamics of single stars contain insight on the formation and evolution of the Galactic halo.

\subsection{Properties of possible evolutionary paths to sdAs}

The sdAs where first unveiled when we mined the SDSS DR12 for pre-ELMs and ELMs, as described in \citet{dr12cat}. They were believed to belong to either of these classes because of the $\log g$ estimated from their SDSS spectra. ELMs show $\log g$ in the range $5.0 \leq \log g \leq 7.0$ and $T\eff \leq 18\,000-20\,000$ \citep[e.g.][]{elmsurveyVII}, filling in the region between the main sequence and the white dwarfs resulting from single evolution in a $T\eff-\log g$ diagram. However, they also show other particular properties. While their colours might be similar to main sequence stars, ELM radii are at least ten times smaller, so they are significantly less luminous than main sequence stars, and thus need to be nearer to be detected at same magnitude. As a consequence, they show higher proper motions than main sequence stars with similar properties. Moreover, they are expected to be encountered still with the close binary companion which led to their mass loss. Most will merge within a Hubble time \citep{brown2016}, implying high radial velocity variation \citep[][e.g., found a median semi-amplitude of 220~km/s]{elmsurveyVII} and somewhat low orbital periods, usually shorter than one day \citep{elmsurveyVII}.

The properties of the precursors of the ELMs, the pre-ELMs, are more difficult to establish, as they have not reached the white dwarf cooling branch yet. If the time-scale for mass loss from the white dwarf progenitor is longer than the thermal time-scale, a thick layer of hydrogen will be surrounding the degenerate helium core. This can lead to residual p--p chain reaction H nuclear burning which can last for several million years \citep[e.g.][]{maxted2014}. Moreover, instead of a smooth transition from pre-ELM to ELM, the star can undergo episodes of unstable CNO burning, or shell flashes. These flashes can shorten the cooling time-scale, by reducing the hydrogen mass on the surface. \citet{althaus2013} find that they occur when $M\leq0.18~M\subsun$, while \citet{istrate2016} find that the minimum mass at which flashes will occur depends on the metallicity of the progenitor. Importantly, these flashes significant alter the radius and effective temperature of a pre-ELM, making it very difficult to distinguish them from main sequence or even giant branch stars. \citet{pietrzynski2012}, for example, found a $0.26~M\subsun$ pre-ELM showing RR~Lyrae-type pulsations --- the flashes caused the object to reach the RR~Lyrae instability strip. Its identification was possible because the system is eclipsing, with an orbital period of 15.2~days, which allowed for an estimate of the mass. \citet{greenstein1973} and \citet{shonberner1978} discuss an interesting example of a post-common envelope binary mimicking a main sequence B star. So pre-ELMs can show $\log g$, $T\eff$ and colours in the same range as main sequence or even giant stars, being even as bright as them. Their ages are more consistent with the halo population than single main sequence stars of similar properties though, since they are at a later stage of evolution.

If found in the halo without a close binary companion inducing enhanced mass loss, a sdA could also be explained as a metal-poor star of type A--F. This is the explanation suggested by \citet{brown2017}. They have, however, based this conclusion on the fact that their fit of pure hydrogen models to metal abundant models seemed to indicate an overestimate in $\log g$ of about 1~dex. As we will show, the change in $\log g$ with the addition of metals to the modelled spectra is actually \textit{not} a constant. Moreover, they have overlooked the possibility that the sdAs are pre-ELMs, which do show $\log g$ in the same range as main sequence stars, but are older and thus should be found in abundance in the halo, whose age is over 10~Gyr --- close to 10 times the expected lifetime in the main sequence of A stars. Main sequence A stars in the halo can only be explained as blue-stragglers, where mass transfer from a companion extends their life in the main sequence, or by rare events of star formation induced by matter accreted to the Galaxy \citep{lance1988,camargo2015}. Main sequence F stars might still be approaching the turn-off point in the halo, so these and other late-type main sequence stars could explain cooler sdAs ($T\eff \lesssim 8000$~K). A key-way to analyse the feasibility of this scenario is analysing the spacial velocities of the sdAs given a main sequence radius, as we will show in Section \ref{distance}.

Finally, another possibility that might explain some sdA is that they are binaries of a hot subdwarf with a main sequence companion of type F, G or K, as found by \citet{barlow2012}. In this kind of binary systems, the flux contribution of both components is similar, so the spectra appear to show only one object, with the lines of the main sequence star broadened by the presence of the subdwarf, explaining the higher values of $\log g$ obtained. However, due to the presence of the subdwarf, which shows $T\eff \geq 20\,000$~K, a higher flux contribution on the UV is expected when compared to main sequence or ELM stars, allowing for telling these objects apart.

In summary, the sdAs physical properties are consistent with basically four different possibilities: (a) pre-ELMs or ELMs; (b) blue-stragglers; (c) metal-poor late-type main sequence stars; (d) hot subdwarf plus main sequence F, G, K binary. Estimated $\log g$ and $T\eff$ should be similar between all possibilities. Colours are similar for pre-ELMs, ELMs and metal-poor main sequence stars, but hot subdwarfs with a main sequence companion should have higher UV flux. ELMs and pre-ELMs should show a close binary companion leading to high radial velocity variations and orbital periods lower than 36~h, according to \citet{elmsurveyVII}, or up to 100~days, according to \citet{sun2017}. Main sequence binaries showing physical parameters in the sdA range, on the other hand, should have periods above $\sim 9$~h \citep{brown2017}. ELMs and pre-ELMs have long evolutionary periods, so they can be detected with ages above 10~Gyr, while main sequence stars of  A-type have main sequence life times lower than 1.5~Gyr, although a companion might delay the evolution by transferring mass as occurs for blue straggler stars.

\section{Data selection}

We have selected all spectra in the SDSS data release 12 \citep[DR12,][]{dr12} containing O, B, A, or WD in their classification with signal-to-noise ratio (S/N) at the $g$ filter larger than 15. This resulted in 56\,262 spectra. They were first fitted with spectral models derived from pure hydrogen atmosphere models calculated using an updated version of the code described in \citet{koester2010}. Objects with $\log g \geq 5.5$ were published in the SDSS DR12 white dwarf catalogue by \citet{dr12cat} and were the first to be called sdAs. Motivated by the fact that many objects showed metal lines in their spectra, we have calculated a new grid with metals added in solar abundances. The grid, whose physical input is discussed in Section \ref{models}, covers $6\,000~\textrm{K} \leq T\eff \leq 40\,000$~K and $3.5 \leq \log g \leq 8.0$. With this grid, we were able to obtain a good fit to 39\,756 spectra out of the initial sample. The remaining objects were mostly close to the border of the grid, either in $T\eff$ or in $\log g$, and are probably giant stars. 723 objects fitted $T\eff > 20\,000$~K and were excluded --- 49 show $\log g > 6.5$ and are canonical mass white dwarfs, while 674 show $\log g <6.5$ and are most likely hot subdwarfs. All the white dwarfs are known with the exception of two new DA white dwarfs (SDSS~J152959.39+482242.4 and SDSS~J223354.70+054706.6). Only 66 out of the 674 possible sdBs are not in the catalogue of \citet{geier2017}; they are listed in Table \ref{sdBs} in the Appendix, with the exception of SDSS~J112711.70+325229.5, a known white dwarf with a composite spectrum that compromised the fit \citep{dr7cat}.

To choose between hot and cool solutions with similar $\chi ^2$, that arise given to the fact that these solutions give similar equivalent width for the lines, we have relied on SDSS \textit{ugriz} photometry, and GALEX \textit{fuv} and \textit{nuv} magnitudes when available. Specifically, we have chosen the solution which gave a $T\eff$ consistent with the one obtained from a fit to the spectral energy distributions (SED) when the $\log g$ was fixed at 4.5. Full reddening correction was applied following \citet{schlegel1998} for the SDSS $ugriz$ magnitudes. For GALEX magnitudes, extinction correction was applied using the $E(B-V)$ value given in the GALEX catalogue, which was derived from the dust maps of \citet{schlegel1998}, and the relative extinction of \citet{yuan2013}, $R_{fuv} =4.89$ and $R_{nuv} = 7.24$, given that such values are not present in the catalogue of \citet{schlegel1998}. \citet{yuan2013} caution, however, that the FUV and NUV coefficients have relatively large measurement uncertainties.

Next, we removed from the sample contaminations from other SDSS pipeline possible classifications that contained our keywords, such as G0V{\bf a}, F8I{\bf b}v{\bf a}r and Calcium{\bf WD}. Those were only 182 objects, leaving a sample of 38\,850 narrow hydrogen line objects with a good solar abundance fit and $T\eff < 20\,000$~K. This sample will be referred to as \textit{sample A} throughout the text.

When we rely on spacial velocity estimates to analyse our sample, only objects with a reliable proper motion are taken into account. Unfortunately, our objects are too faint to be featured in the DR1 of Gaia, so we used the proper motions of \citet{tian2017}, which combine Gaia DR1, Pan-STARRS1, SDSS and 2MASS astrometry to obtain proper motions. To flag a proper motion as good, we required that the  distance to nearest neighbour with $g>22.0$ was larger than 5'', that the proper motion was at least three times larger than its uncertainty, and that the reduced $\chi$-squared from the evaluation of proper motions in right ascension and declination was smaller than 5.0. This left 16\,656 objects with a reliable proper motion, with an average uncertainty of 2.0~mas/yr, to be referred to as \textit{sample B} in the text.

In order to estimate the contamination by outliers, we have compared GPS1 proper motions to both the Hot Stuff for One Year \citep[HSOY,][]{hsoy} catalogue and the catalogues by \citet{munn2004} and \cite{munn2014}, directly available at the SDSS tables. HSOY combines positions from Gaia DR1 and the PPXML catalogue \citep{ppmxl}, while Munn et al. combine SDSS and USNO-B data. Hence they are not completely independent, but nevertheless useful to find possible outliers. We find only 69 objects whose proper motions differ by more than 3-$\sigma$ when comparing GPS1 and HSOY, and 110 objects when comparing to Munn et al. They represent less than 1 per cent of the sample, so we decided to keep them as part of sample B, since it does not affect the analysis, and GPS1 is regarded as the best proper motion catalogue available for our objects.

\section{Spectral analysis}

\subsection{The models}
\label{models}

The code to calculate atmospheric models and synthetic spectra was originally developed for white dwarfs \citep{koester2010}. In that area it has been used and tested for decades and proven to be reliable \citep[e.g.][]{dr7cat,dr10cat,dr12cat}. It has also been used successfully for sdB stars with surface gravity around $\log g = 5.5$ \citep[e.g.][]{dr12cat}. The identification of sdA stars was a by-product of our study of white dwarfs in the SDSS DR12. We extended our grid to lower surface gravity in order to better identify and separate these sdAs from DAs and sdB. This extension is not meant to be used for a full-fledged analysis of main sequence stars. It rather serves as an indicator of the luminosity class, given the external uncertainties of 5--10 per cent in $T\eff$ and 0.25~dex in $\log g$.

To estimate internal uncertainties, we have compared our estimates for objects with duplicate spectra. We have found the average difference to be 0.55 per cent in $T\eff$, with a standard deviation of 2.8 per cent, and of 0.047~dex in $\log g$, with a standard deviation of $0.133$. Interestingly, these values are not significantly dependent on $S/N$ for $S/N > 15$. We have also found no variation in these internal uncertainties when excluding objects cooler than $8\,000$~K from the sample. Hence the behaviour of the internal uncertainty does not seem to depend on either $T\eff$ or $S/N$ for the ranges considered here. The external uncertainty is higher for lower $T\eff$, due to the decreasing strength of the lines, making them less sensitive as gravity indicators. That is, however, hard to quantify.

The models include metals up to $Z = 30$ with solar abundances in the equation of state, and include also the H$_2$ molecules. This ensures that the number densities of neutral and ionized particles are reasonable, which is important for the line broadening, in particular the Balmer lines. The tables of \citet{tremblay2009}, which include nonideal effects, are used to describe the Stark broadening of the Balmer lines. The occupation probability formalism of \citet{hummer1988} is taken into account for all levels of all elements. Absorption from metals is not included. We have tested that the addition of the photoionization cross sections of metals with the highest abundances does not result in significant changes in the A star region. Line blanketing for the atmospheric structure uses only the hydrogen lines. The synthetic spectra, however, include approximately the 2400 strongest lines of all elements included in the range 1500--10000~\AA.

\begin{figure*}
  \includegraphics[height=\textwidth,angle=-90]{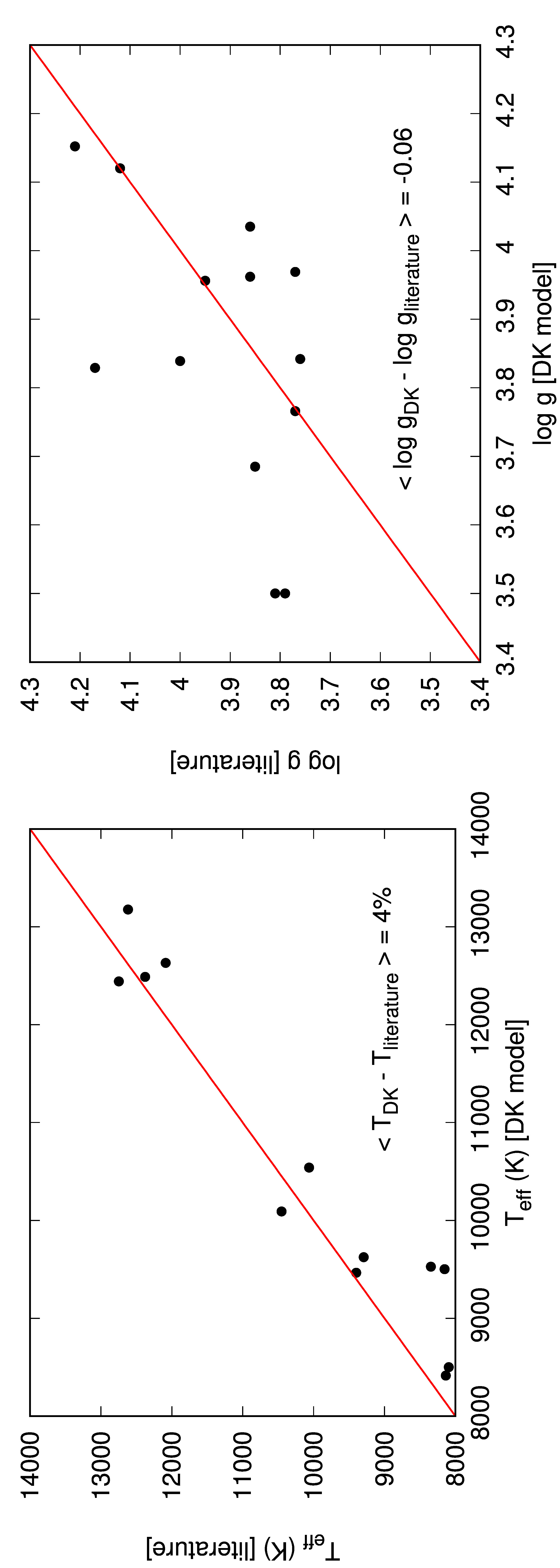}
  \caption{Left panel shows the comparison between the $T\eff$ obtained with our models (DK), and values from the literature. Right panel shows the same comparison for $\log g$. The average difference is of about 4 per cent in $T\eff$, and of -0.06 in $\log g$, which is even lower than the external error. The dashed red line in both panels shows where both determinations would be equal.}
  \label{comp}
\end{figure*}

As a test for the validity of the models we have used a similar setup as for our sdA/ELM spectral fitting for a selection of known A stars. These include some of the objects with $\log g > 3.75$ from \citet{allendeprieto2016}, as well as Vega from \citet{bohlin2007}. As Fig. \ref{comp} shows, the average differences between our obtained values and the values from the literature are, in average, 4~per cent in $T\eff$ and -0.06~dex in $\log g$, with no great discrepancies or systematic differences. This average difference can be easily explained by the dominant external errors. In Appendix \ref{more_models}, we compare our estimates to those of the SDSS pipelines, whose grids have much smaller coverage than our own.

\subsection{Spectral fits}

\begin{figure}
  \includegraphics[width=\columnwidth]{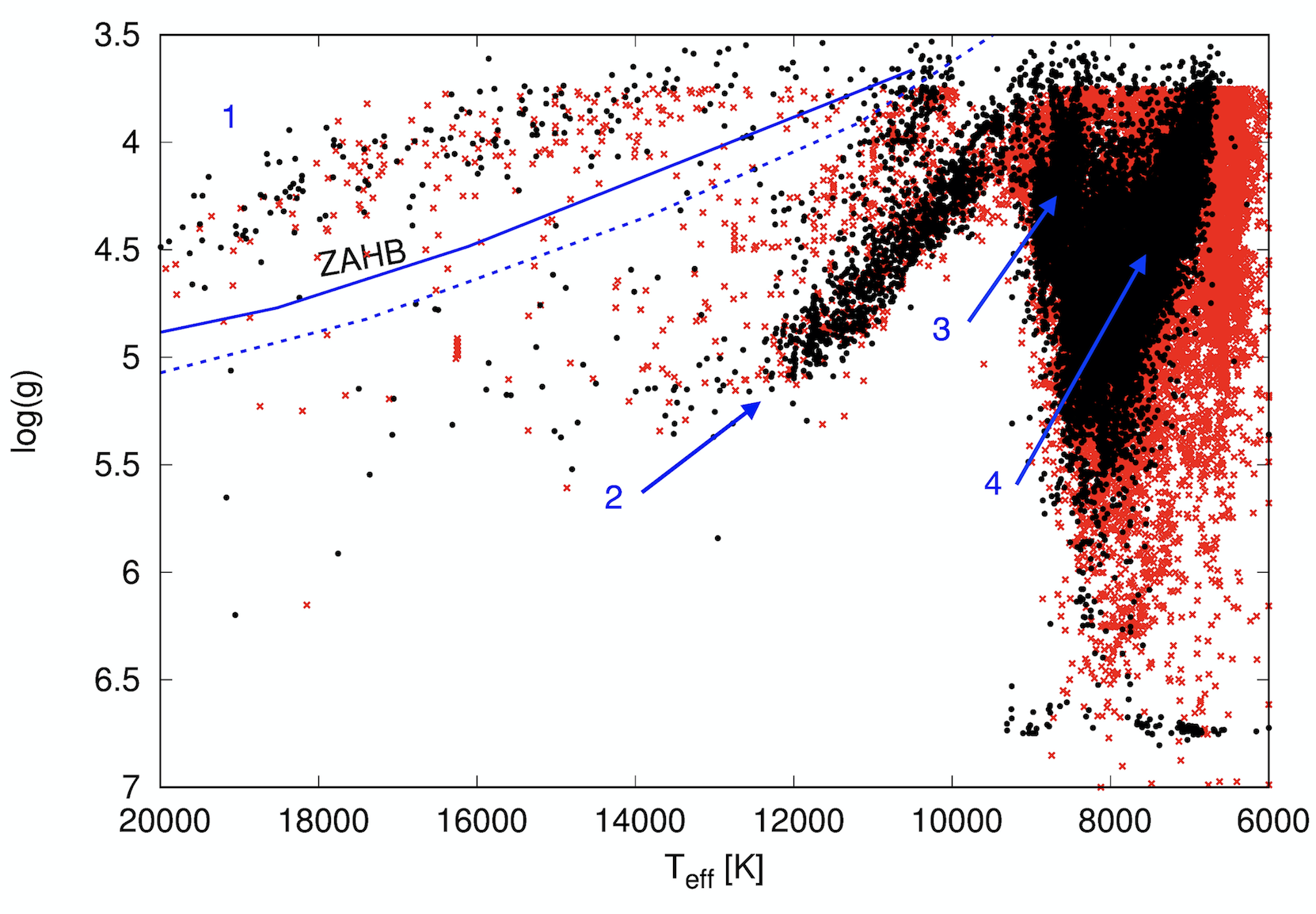}
  \caption{$T\eff - \log g$ diagram showing the results of our pure hydrogen spectral fits as red crosses, and the updated result with metals added in solar abundance as black dots. The two distributions are shifted due to the changes in $T\eff$ and $\log g$ for individual objects. The continuous blue line indicates the ZAHB, above which stars might be burning He in the core. Its position depends on metallicity; the continuous line assumes $Z = 0.0001$, the dashed line is for solar metallicity. Different sequences are labelled as 1, 2, 3, and 4. They reflect different temperature regimes described in the text.}
  \label{fit_arrows}
\end{figure}

Fig. \ref{fit_arrows} shows the $T\eff - \log g$ diagram with the values obtained from the two different models, pure hydrogen and solar abundance, for objects with good fit in both cases. It can be noted that the distribution shifts as a whole with the addition of metals to the models. Four sequences can be distinguished. At the hot low gravity end, some objects (labelled as sequence 1 in Fig. \ref{fit_arrows}) are above the ZAHB; they could hence be blue horizontal branch stars. They are kept in the sample because, as we will show later, this region of the diagram can also be reached through binary evolution. There are a few hot objects between 10\,000--12\,000~K (sequence 2), and the bulk of the distribution is between 7\,000--10\,000~K. Careful inspection, especially at the low $\log g$ end, suggests this region can also be split in two regimes around 8\,000~K (sequences 3 and 4).

\begin{figure}
  \includegraphics[height=\columnwidth,angle=-90]{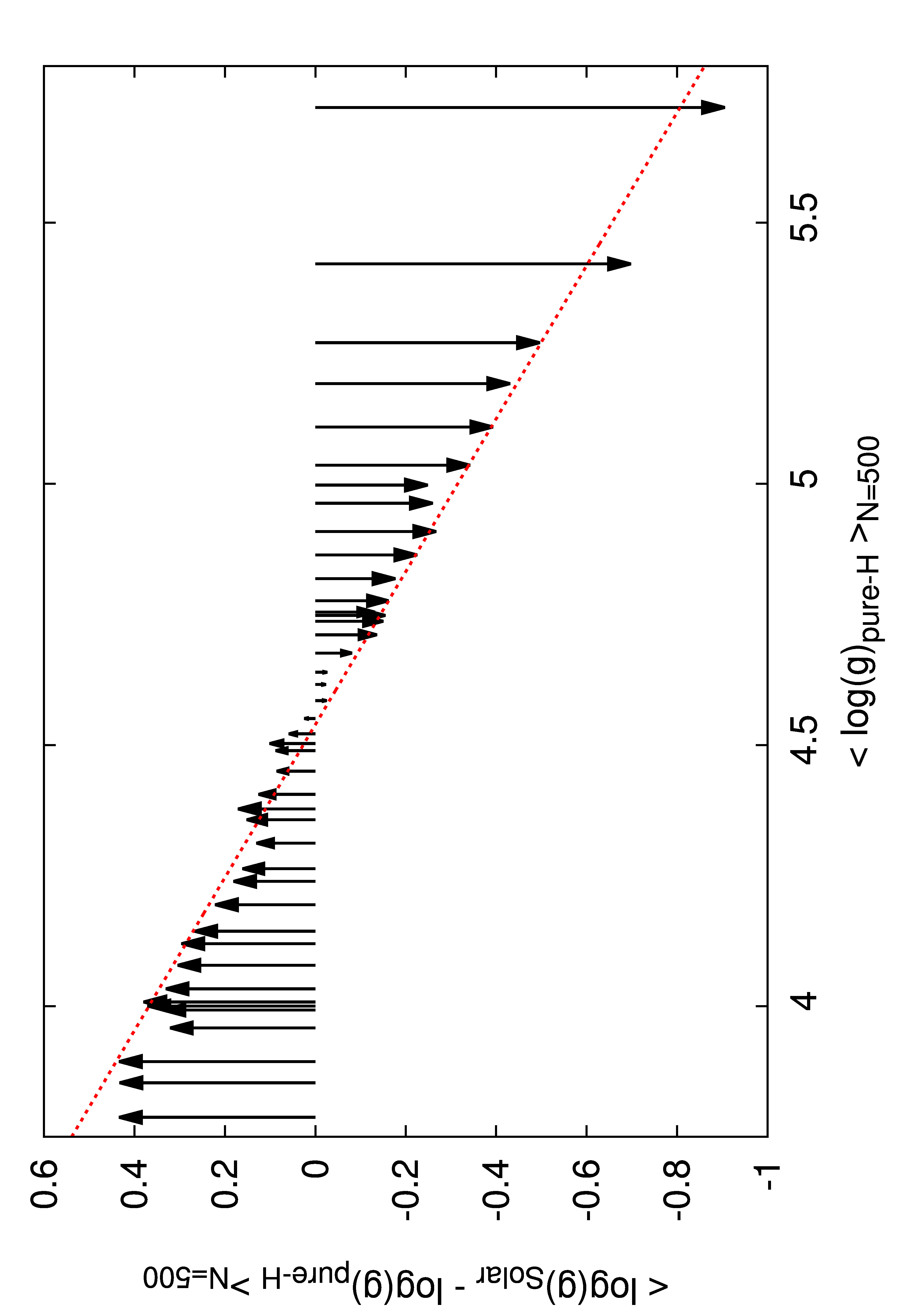}
  \captionsetup{singlelinecheck=off}
  \caption[ ]{Shift in $\log g$ with the addition of metals in solar abundances as a function of the $\log g$ given by the pure-H models. Values were averaged over 500 objects sorted by $\log g$. The shifts are well described by a linear fit 
    \begin{displaymath}
      \Delta \log g = -0.68(0.01)\,\log g_{\textrm{pure-H}} + 3.10(0.06)
    \end{displaymath}
    with the pure-H values being overestimated by almost 1.0~dex above $\log g = 5.5$. This is a similar result to the obtained by \citet{brown2017} when fitting pure hydrogen model to synthetic main-sequence spectra.}    
  \label{arrows_logg}
\end{figure}

We have analysed the change in $\log g$ for objects in sample A as a function of the pure hydrogen $T\eff$ and $\log g$ values. We took into account only objects whose two estimated $T\eff$ values differed by less than 500~K. We found a clear trend when plotting $\log g_{\textrm{Solar}} - \log g_{\textrm{pure-H}}$ as a function of $\log g_{\textrm{pure-H}}$, as can be seen in Fig. \ref{arrows_logg}. The larger the $\log g_{\textrm{pure-H}}$, the smaller is the $\log g$ obtained from solar abundance models compared to pure hydrogen models. The shift, however, is not a constant value as suggested by \citet{brown2017}, but it rather behaves as a linear function of  $\log g_{\textrm{pure-H}}$. Above $\log g \sim 5.5$, the solar abundance $\log g$ is in fact about 1~dex smaller than the value obtained from pure hydrogen models, as they have found. This explains their conclusion and that of \citet{hermes2017} when analysing the $\log g \geq 5.5$ DR12 sdAs: they are exactly in this range where the difference between these two values is maximal. However, it is important to emphasise that the solar abundance model is not necessarily the correct one; while many sdAs do show clear signs of metals in their spectra, others seem to be almost free of metals.

\begin{figure}
 \includegraphics[height=\columnwidth,angle=-90]{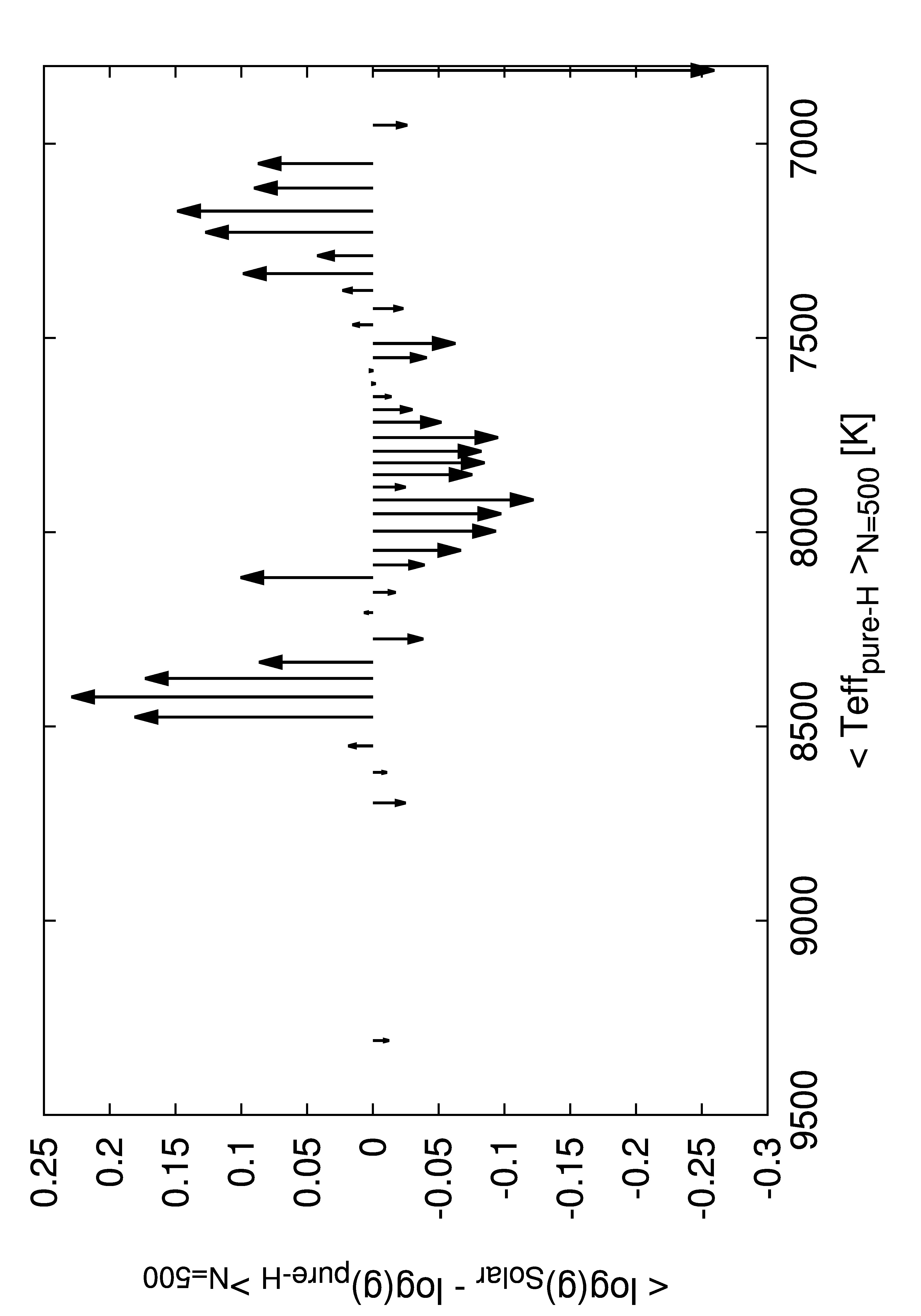}
  \caption{Change in $\log g$ when metals were added to the models as a function of the effective temperature of the pure-H models. The $T\eff$ and the change in $\log g$ were averaged over 500 objects, sorted by $T\eff$. The systematic effect found as a function of $\log g_{\textrm{pure-H}}$ implies on a correlation also in $T\eff$, depending on how each range of $\log g$ is sampled in each bin of $T\eff$. Around 7\,000~K, for example, most objects have $\log~g < 4.5$, where the shift in $\log g$ points upwards in Fig. \ref{arrows_logg}, what is also seen here.}
  \label{arrows_Teff}
\end{figure}

This systematic trend also reflects on the dependance of the $\log g$ change with $T\eff$, shown in Fig. \ref{arrows_Teff}. At $T\eff \sim 8\,500$~K, there are objects spanning all the $\log g$ range (sequence 3 in Fig. \ref{fit_arrows}), but a prevalence of objects with lower $\log g$, which have an upward correction. Hence the same upward correction is seen in this $T\eff$ range. Between $7\,500 - 8\,000$~K, a gap in the lower $\log g$ objects can be seen in Fig. \ref{fit_arrows}, which moves the correction downwards. Finally, below $T\eff \sim 7\,500$~K (sequence 4), most objects show $\log g \leq 4.5$, so the correction moves upwards again. Close to the cool border of $T\eff$, most objects are also close to the lower border in $\log g$, which is 3.75 for the pure-hydrogen models and 3.5 for the solar abundance models, implying on an average difference of 0.25. There can of course be differences in metallicity and errors in the determination, so individual objects can somewhat obscure these trends. 

The solar abundance solutions put most of the 2\,443 sdAs published in by \citet{dr12cat} in the main sequence range, with the exception of 39 objects which still show $\log g \geq 5.0$. Only seven out of those maintain $\log g \geq 5.5$ in the solar abundance models. It is important to notice, however, that these higher values of $\log g$ can rise from statistics alone given an external uncertainty of about 0.25 even if the correct $\log g$ for these objects is about 4.5, so these objects should be analysed with caution.

Two of the $\log g > 5.0$ objects were published in the ELM Survey, SDSS~J074615.83+392203.1 \citep{elmsurveyIII} and SDSS~J091709.55+463821.7 \citep{elmsurveyVI}. SDSS~J0746+3922 was not confirmed as a binary; the published solution of $T\eff = 12\,130 \pm 400$~K and $\log g = 5.98 \pm 0.12$ agrees in $\log g$ with our pure hydrogen solution, $T\eff = 8300$~K and $\log g = 5.85$, but there is a discrepancy in $T\eff$. The UV colours favour the hotter solution. Our solar metallicity solution gives a slightly lower $\log g$ of $5.481 \pm 0.017$ and $T\eff = 8326 \pm 9$~K\footnote{Quoted uncertainties in our values of $T\eff$ and $\log g$ are formal fit errors. The external uncertainties in the models are much larger, as discussed in Section \ref{models}.}. SDSS~J0917+4638 was confirmed as a binary with period of 7.6~h and amplitude of 150~km/s. The solution published in \citet{elmsurveyVI}, $T\eff = 12240 \pm 180$ and $\log g = 5.75 \pm 0.04$, agrees with our solar metallicity values of $T\eff = 12958 \pm 111$ and $\log g =  5.842 \pm 0.029$. Our pure hydrogen fit indicates $T\eff = 9\,600$~K and $\log g = 5.00$.

Another object which maintained $\log g > 5.0$ is SDSS~J075017.35+400441.2, an eclipsing binary analysed in \citet{brown2017}. Our solar metallicity fit gives $T\eff = 8071 \pm 15$~K and $\log g = 5.019 \pm 0.038$, a $\log g$ significantly lower than the pure hydrogen value of 5.619. \citet{brown2017} points out that the SEGUE stellar parameter pipeline (SSPP) gives a much lower $\log g$ of $4.229 \pm 0.155$. However, the SSPP grid has no model above $\log g = 5.0$. The obtained period from their radial velocity orbital fit agrees with the photometric period of 28~h. They obtain a radial velocity amplitude of 36.2~km/s and conclude the star is best explained by a metal-poor main-sequence binary, which is consistent with our solar metallicity solution given the external uncertainties. The star's detected proper motion is not significant ($2.03 \pm 1.96$~mas/yr), but the distance obtained from its distance modulus is over 14~kpc --- in the Galactic halo. If indeed a main sequence A star, it can only be explained as blue straggler whose main sequence lifetime was significantly extended due to mass accreted from the companion.

SDSS~J014442.66-003741.7, which was classified as $\delta$-Scuti by \citet{bhatti2010} given its Stripe 82 SDSS data, also has $\log g > 5.0$ in both our models. \citet{bhatti2010} obtained a period of 1.5~h, which could also be explained as a $g$-mode pulsation of a pre-ELM star, given our solar metallicity fit of $T\eff = 7\,949 \pm 35$ and $\log g = 5.18 \pm 0.11$. The object's proper motion in the GPS1 proper motion table, $8.48 \pm 3.59$~mas/yr has too high uncertainty to allow any further conclusions on the object's nature. The object is relatively faint, $g=19.8$, so the SDSS subspectra have too low SNR to allow good estimates of radial velocity. Better data are needed in order to establish the nature of this star. As discussed in S\'{a}nchez-Arias et al. (\textit{submitted to A\&A}), period spacing and rate of period change can be used to tell pre-ELMs and $\delta$-Scuti stars apart.

Given that the changes in $\log g$ and $T\eff$ can go up or down, many other objects are raised above the main sequence $\log g$ limit. Table \ref{Tlog_table} lists the 408 objects with $5.5 \leq \log g < 7.0$ and $7\,000 \leq T\eff \leq 20\,000$~K. Other 82 objects with $\log g$ in this range but $T\eff < 7\,000$~K are omitted because we believe our models to be unreliable below this limit. The general weakness of the lines and the uncertainty of neutral line broadening at these temperatures make the $\log g$ difficult to estimate.

\begin{table*}
  \caption{Obtained values of $T\eff$ and $\log g$ for the 408 analysed SDSS objects in the range $5.5 \leq \log g < 7.0$ and $7\,000 \leq T\eff \leq 20\,000$~K. The quoted uncertainties are formal fitting errors; external uncertainties in the models are 5--10 per cent in $T\eff$ and 0.25~dex in $\log g$. The plate-modified Julian date-fibre, or P-M-F, identifies the SDSS spectrum for the object from which the solution was obtained. The S/N of the spectrum is given at the $g$-band. The full table can be found in the on-line version of this paper.}
  \label{Tlog_table}
  \begin{tabular}{ccccc}
    \hline
    SDSS~J & P-M-F & S/N$_g$ & $T\eff$ (K) & $\log g$ \\
    \hline
    112616.66-010140.7 & 0281-51614-0243 & 14.72 &  8312(46) & 5.624(0.170) \\ 
    113704.83+011203.6 & 0282-51658-0565 & 44.29 &  8122(11) & 5.544(0.040) \\ 
    130149.63+003823.8 & 0293-51689-0581 & 19.07 &  7150(37) & 6.724(0.004) \\ 
    130717.12-002639.4 & 0294-51986-0174 & 21.25 &  8167(36) & 5.572(0.134) \\ 
    134428.86+002820.4 & 0300-51666-0342 & 51.18 &  7246(11) & 6.704(0.001) \\ 
    135042.43-002004.7 & 0300-51943-0102 & 17.84 &  8582(36) & 5.606(0.150) \\ 
    140114.04-003553.6 & 0301-51641-0104 & 16.31 &  7403(40) & 6.747(0.006) \\ 
    140126.86+003156.3 & 0301-51641-0584 & 14.50 &  8040(41) & 5.728(0.135) \\ 
    121715.08-000928.3 & 0324-51666-0039 & 16.94 &  8358(38) & 5.533(0.142) \\ 
    162220.66-002000.4 & 0364-52000-0272 & 35.80 &  8242(20) & 5.672(0.082) \\ 
    \hline
  \end{tabular}
\end{table*}

\section{Colours}

While spectra are considered the most reliable way to estimate the physical properties of a star, the colours of an object alone can still tell us something about its nature and be used as a complement to spectral results, especially when the colors include the ultraviolet region, not included in most spectra. \citet{girven2011}, for example, used colour-colour cuts in the SDSS photometry to identify DA white dwarfs with an efficiency of returning a true DA of 62 per cent, obtaining a 95 per cent complete sample. This type of approach was also used by \citet{dr7cat}, \citet{dr10cat}, and \citet{dr12cat} to select white dwarf candidates in the SDSS database. This method relies notably on the $(u-g) \times (g-r)$ diagram, where models are significantly dependent on $\log g$. However, for pre-ELMs and ELMs, the lower $\log g$ gives them colours more similar to main sequence stars, making this method less effective.

\begin{figure}
	\includegraphics[width=\columnwidth]{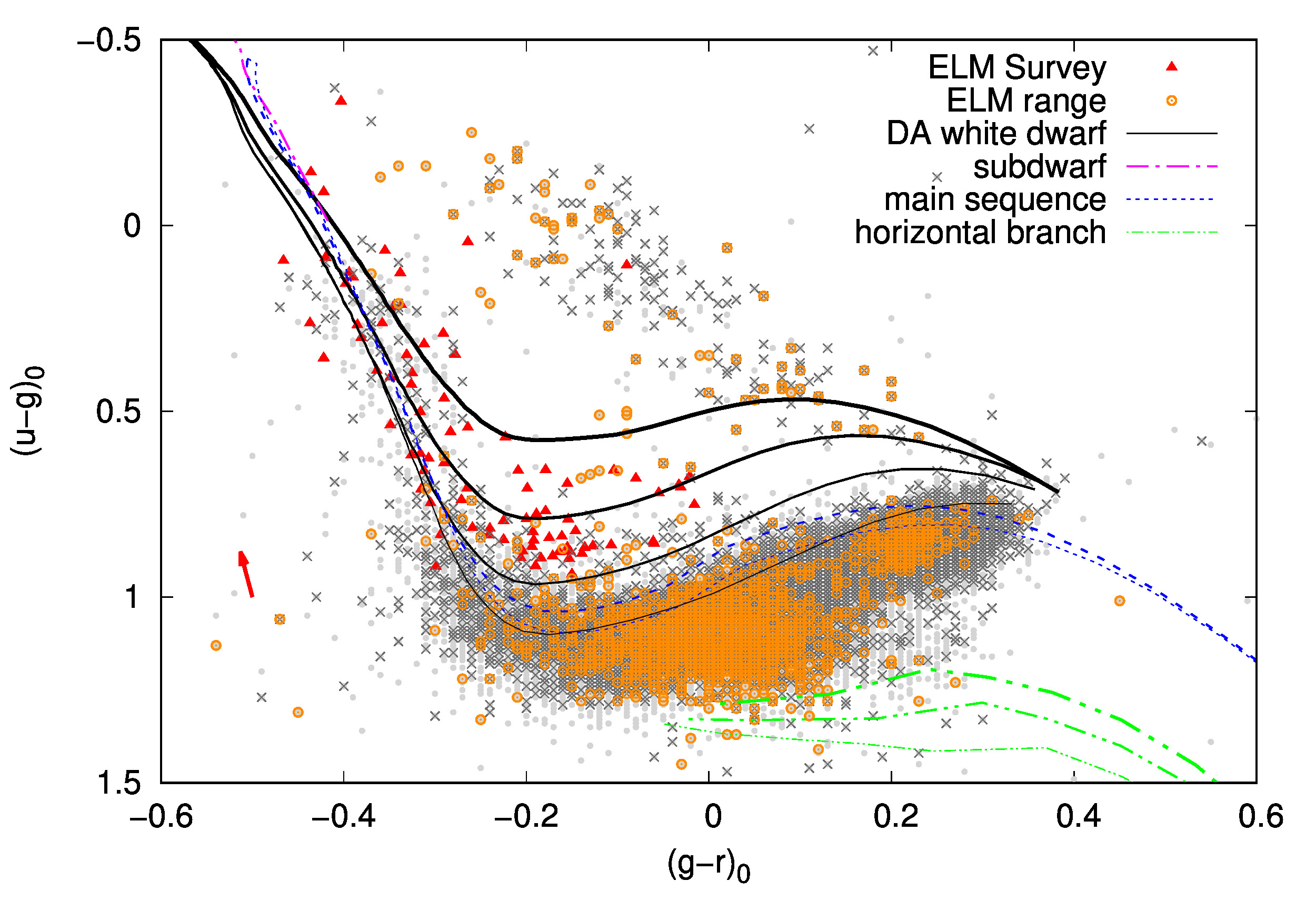}
        \caption{Colour-colour diagram showing sample A as dots in light grey, sample B as dark grey crosses, and the known ELMs from \citet{elmsurveyVII} as red triangles. Objects whose obtained spectral fit places them in the ELM range $T\eff \leq 20\,000$~K and $5.0 \leq \log g \leq 7.0$ are marked with orange circles. The red arrow indicates the average vector of the reddening correction. Some theoretical models are included to guide the eye; the increasing thickness of the lines reflects an increasing $\log g$. The DA white dwarf models in black are obtained from our pure-hydrogen spectral models by convolving them with the SDSS filters. They span $\log g$ 4.0--7.0 in steps of 1.0 from bottom to top. Subdwarf, main sequence, and horizontal branch models are from \citet{lenz1998}. The subdwarf model assumes $\log g = 5.00$ and $[ M/H ] = 0.0$, and covers $20\,000$~K$\leq T\eff \leq 50\,000$~K. The selected main sequence models have fixed $[ M/H ] = -5.0$, with $\log g = 4.0, 4.5$ and $4\,250$~K$\leq T\eff \leq 40\,000$~K. Finally, horizontal branch models have $[ M/H ] = -1.0$, $\log g = 1.0, 1.5, 2.0$ and span $3\,500$~K$\leq T\eff \leq 26\,000$~K. }
    \label{ugr}
\end{figure}

In Fig. \ref{ugr} we show the $(u-g)_0 \times (g-r)_0$ diagram for samples A and B. Full reddening correction is applied following \citet{schlegel1998}. For comparison, we also show the confirmed ELMs from the ELM Survey as published in \citet{elmsurveyVII}. Objects that were placed in the ELM range, $T\eff \leq 20\,000$~K and $5.0 \leq \log g \leq 7.0$, when fitted with our solar metallicity models, are marked with orange circles. They could be interpreted as extending the ELM strip to cooler temperature, but remarkably most of them lie below the $\log g = 5.0$ model line in this colour-colour diagram, despite the fact that spectroscopy indicates $\log g > 5.0$. This might suggest that there is still some missing physics in our spectral models: the addition of metals alone does not solve the discrepancy. Possibly some opacity included in the models needs better calculations, as might the case for broadening of the Balmer lines by neutral hydrogen atoms. He contamination through deep convection may also play a role. A possibility that cannot be discarded is that the extinction correction is not accurate due to reasons such as variations on dust type and size, or the object being within the Galactic disk. 

\begin{figure}
	\includegraphics[width=\columnwidth]{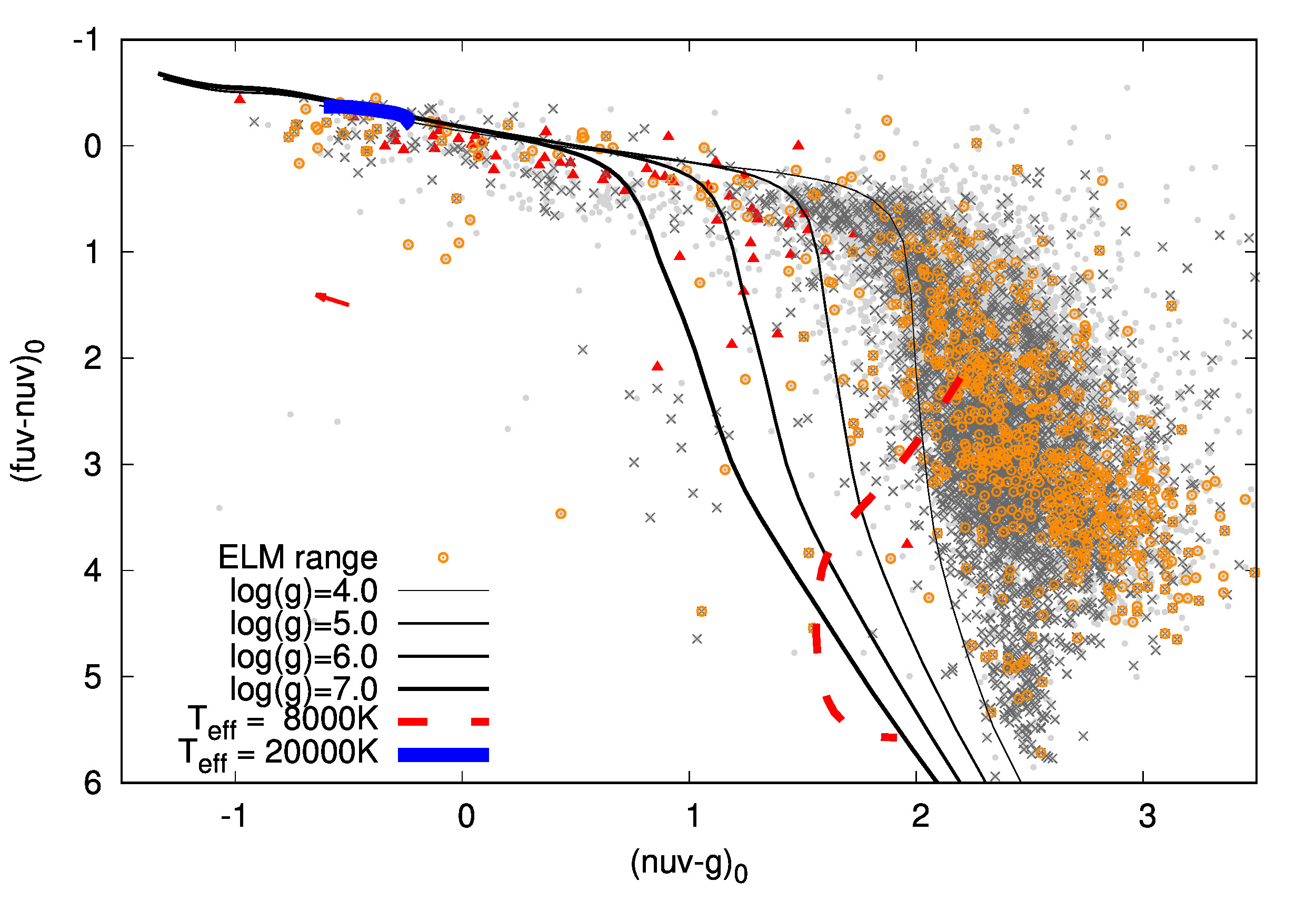}
    \caption{Diagram showing the $(fuv-nuv)_0$ and $(nuv-g)_0$ colours. The colours of each sample are the same as in Fig. \ref{ugr}. The red arrow indicates the average reddening correction vector. The indicated models were obtained from our pure-hydrogen spectral models. Despite the reddening correction, the data still seems dislocated in relation to the models, suggesting this correction might be underestimated.}
    \label{fnuvg}
\end{figure}

We have also analysed the GALEX UV magnitudes, far-ultraviolet ($fuv$) and near-ultraviolet ($nuv$), when available. Fig. \ref{fnuvg} shows a $(fuv-nuv)_0 \times (nuv-g)_0$ diagram for samples A and B; the objects for which we have obtained $T\eff \leq 20\,000$~K and $5.0 \leq \log g \leq 7.0$ are marked as orange circles. Extinction correction was applied using the $E(B-V)$ value given in the GALEX catalogue, $R_{fuv} = 4.89$ and $R_{nuv} = 7.24$ \citep{yuan2013}. The colours suggest again $\log g$ lower than the estimated spectroscopically. However, extinction correction is even more uncertain in the ultraviolet than in the visible region, so again it should not be discarded that the correction is underestimated.

This diagram is especially useful in identifying sdB + FGK binaries, which should have significant flux in the UV due to the hot subdwarf component showing $T\eff \gtrsim 20\,000~K$. In Fig. \ref{fnuvg}, there is a clustering of objects with $(nuv-g)_0 < -0.4$; many of them show radial velocity differences larger than 100~km/s in the SDSS subspectra that compose the final spectrum. About half of the sdBs are found to be in close binary systems \citep[e.g.][]{heber2016}, with many showing radial velocity amplitudes in this range \citep[e.g.][]{copperwheat2011}. This considered, we suggest that sdAs showing $(nuv-g)_0 < -0.4$ ---  about 0.5 per cent --- can be explained as sdB + FGK binaries.

Notably, two published ELMs are in this colour range: SDSS~J234536.46-010204.9 and SDSS~J162542.10+363219.1. SDSS~J2345-0102 was analysed in \citet{elmsurveyII}. They obtained $T\eff = 33\,130 \pm 450$ and $\log g = 7.20 \pm 0.04$ and found no evidence of radial velocity variations, suggesting this object is 0.42~$M\subsun$ white dwarf --- therefore a low-mass white dwarf, which are often found to be single, rather than an ELM. The obtained $T\eff > 20\,000$~K and $\log g > 7.0$ make it easier to distinguish this object from the sdAs, so it is not affected by our $(nuv-g)_0 < -0.4$ criterion. On the other hand, SDSS~J1625+3632 which was also analysed in \citet{elmsurveyII}, has its estimated parameters, $T\eff = 23\,570 \pm 440$ and $\log g = 6.12 \pm 0.03$, close to the range where we put the sdAs. \citet{elmsurveyII} found it to present a small semi-amplitude of $K = 58.4$~km/s and a period of 5.6~h, suggesting it to be a 0.20~$M\subsun$ white dwarf with most likely another white dwarf as a companion. However, they point out that their obtained physical parameters are very similar to the sdB star HD~188112 \citep{heber2003}, and mention that the 4471~\AA{} line, a common feature of sdB stars, can be detected on the spectrum of the object. All this combined suggests that this object is rather a sdB than an ELM, fitting the $(nuv-g)_0 < -0.4$ criterion.

\begin{figure}
	\includegraphics[width=\columnwidth]{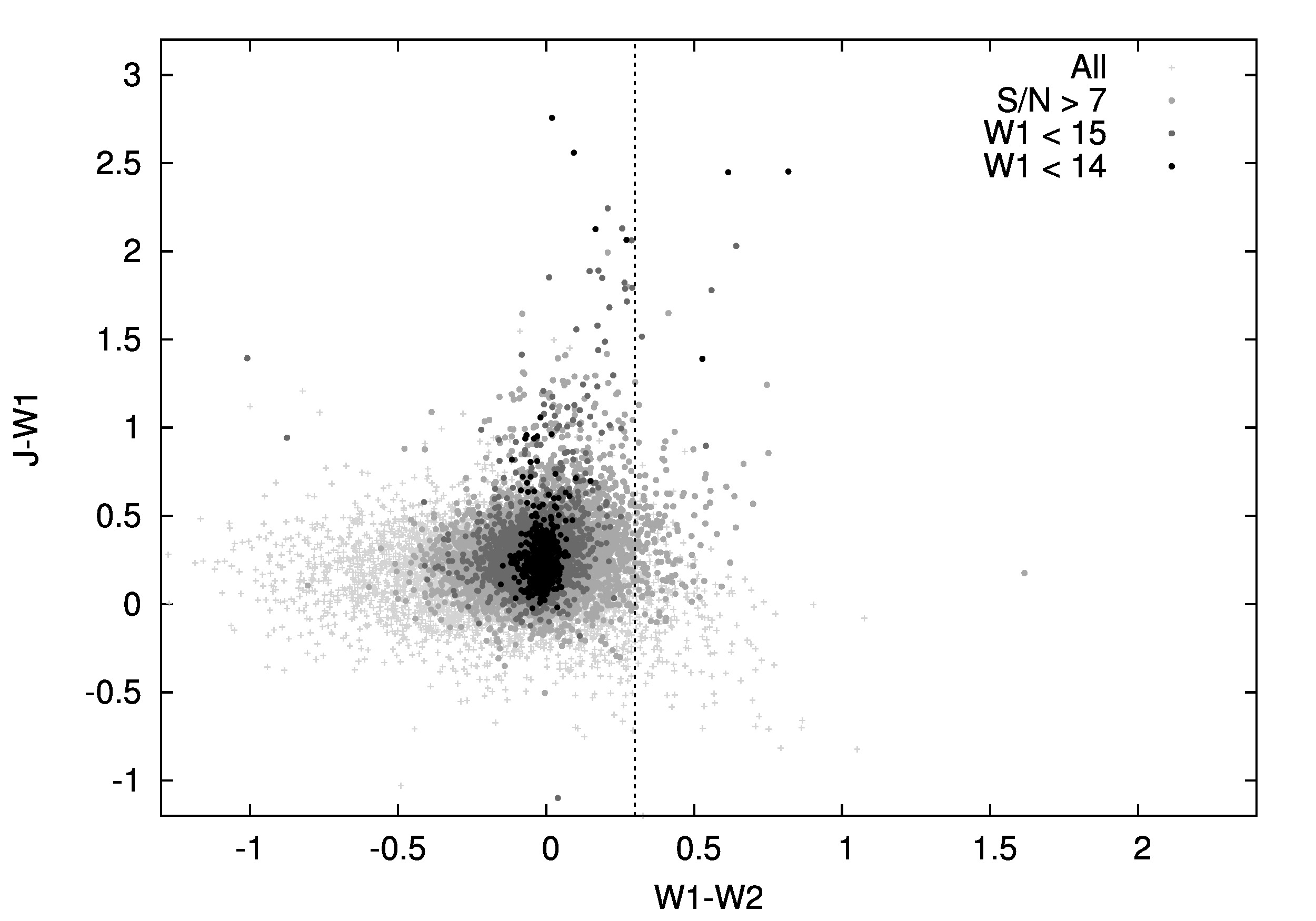}
    \caption{$(J-W1) \times (W1-W2)$ colour-colour diagram for sample A. The whole sample is shown as light grey crosses, objects with $S/N > 7$ at both $W1$ and $W2$ filters are shown as grey dots. The dark grey dots indicate objects brighter than $W1 = 15$, and the black dots are those brighter than $W1 = 14$. The dashed vertical line is the $(W1 - W2) = 0.3$ limit, above which the objects might have infrared excess as suggested by \citet{hoard2013}.}
    \label{wise}
\end{figure}

Finally, we searched for infrared excess due to a cool companion star using data from the Wide-field Infrared Survey \citep[WISE,][]{wise} and the Two Micron All Sky Survey \citep[2MASS,][]{2mass}. We follow the approach of \citet{hoard2013}, who searched for candidate white dwarfs with infrared excess by examining a $(J-W1) \times (W1-W2)$ diagram, suggesting $(W1-W2)>0.3$ as an indication of possible excess. As both white dwarfs and sdAs show hydrogen-dominated spectra, with very few lines in the infrared, the infrared flux in both cases depends basically on $T\eff$, thus the method is suitable for analysing the sdAs. \citet{hoard2013} restrict their analysis to objects with $S/N > 7$ at both W1 and W2. Using this same criterion, we find only about 1.3 per cent of the sample to possibly show infrared excess. The percentage is similar when we consider only objects brighter than $W1 = 14$ or than $W1 = 15$, as illustrated in Fig. \ref{wise}

\section{Distance and motion in the Galaxy}
\label{distance}

\begin{figure}
	\includegraphics[angle=-90,width=\columnwidth]{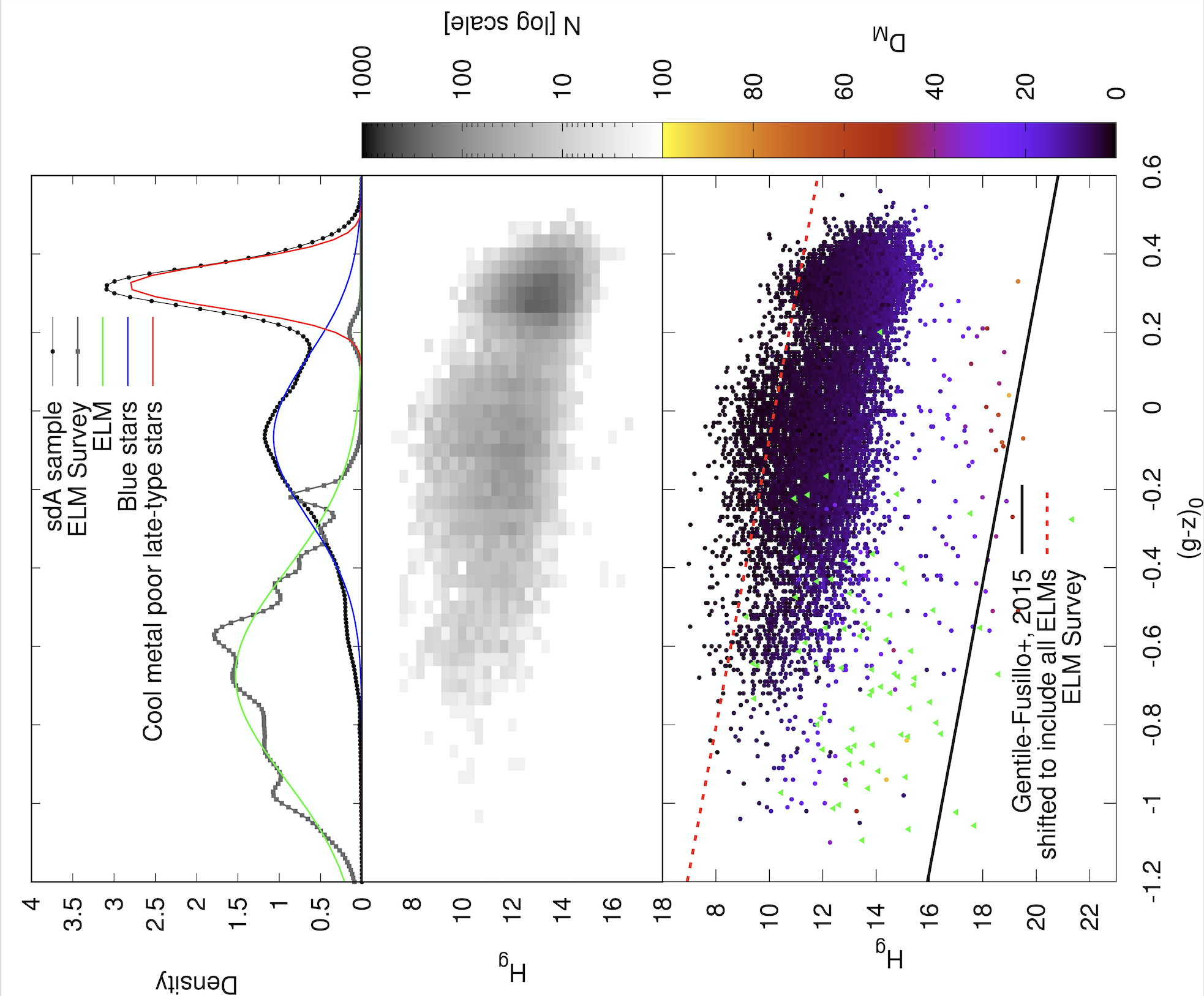}
    \caption{The bottom panel shows the $H_g \times (g-z)_0$ diagram \citep[see e.g.][]{nicola2015}, with the objects in sample B colour coded according to their Mahalonobis distance to the halo given a main sequence radius. Known ELMs are shown as green triangles for comparison. Middle panel shows the same diagram as a bidimensional histogram. The top panel shows the densities assuming each object as a Gaussian to account for the uncertainty; it becomes clear that there are two populations of objects within the sdA sample. The suggested limit for white dwarf detection with probability equal to 1.0 given by \citet{nicola2015} is indicated as a black solid line. Most known ELMs, due to their larger radius implying a smaller reduced proper motion, since they can be detected at larger distances, are not below the white dwarf limit. A reference line, defined arbitrarily shifting the estimate of \citet{nicola2015} to include all known ELMs is shown as a red dashed line. Most sdAs are also below such line.}
    \label{ppm}
\end{figure}

One further step in the separation of white dwarfs from main sequence objects is taking into account measured proper motions \citep[e.g.][]{nicola2015}. As white dwarfs are compact objects, they have smaller radius and therefore are fainter than main sequence stars with same temperature. Due to their degenerate nuclei, white dwarfs have a mass-radius relationship $R \sim M^{-1/3}$, implying that the smaller the mass, the larger the radius. Thus ELMs have larger radius and are brighter than common mass white dwarfs. Still, their radii are of the order of 0.1~$R\subsun$, so they should be about 10 times closer than main sequence stars with similar $T\eff$ to be seen at similar apparent magnitude, showing higher proper motion. The picture is more complicated when the pre-ELMs are considered. Mostly because of the CNO flashes, they can be as bright as main sequence stars, so proper motion cannot be used as a criterion to tell these objects apart. Fig. \ref{ppm} shows a reduced proper motion ($H_g$) versus $(g-z)_0$ diagram. Only sample B, containing objects with reliable proper motion, is taken into account. Here the reduced proper motion is evaluated as
\begin{eqnarray}
H_g &=& g_0 + 5\log(\mu[''/yr]) + 5.
\end{eqnarray}
It can be interpreted as a proxy for the absolute magnitude: the higher the reduced proper motion, the fainter the object.

The objects are colour coded by their Mahalonobis distance $D_M$ \citep[e.g.][]{elmsurveyIV} to the halo when a main sequence radius is assumed. The Mahalonobis distance is given by
\begin{eqnarray}
D_M = \sqrt{ \frac{\left( U - \langle U \rangle \right)^2}{\sigma^2_U} + \frac{\left( V - \langle V \rangle \right)^2}{\sigma^2_V} + \frac{\left( W - \langle W \rangle \right)^2}{\sigma^2_W}},
\label{mahalonobis}
\end{eqnarray}
where we have assumed the values of \citet{kordopatis2011} for the halo mean velocities and dispersions. The Mahalonobis distance measures the distance from the center of the distributions in units of standard deviations; hence considering the size of our sample and assuming a Gaussian behaviour, all objects should show $D_M < 4.0$. Nonetheless, when a main sequence radius is assumed, about 74 per cent of the objects show $D_M > 4.0$. When we assume an ELM radius for these objects, this number falls to less than 2 per cent.

Fig. \ref{ppm} suggests that most of the objects with $T\eff$ and $\log g$ in the ELM range have, in average, $H_g$ lower than the estimated value for known ELMs. This, combined with the fact that they seem to be in the same region of the diagram as the $\log g < 5.0$ objects, could again be seem as an indication of missing physics in the models leading to an overestimate of the $\log g$. However, their reduced proper motion is consistent with a tentative limit based on \citet{nicola2015}, but including all ELMs. This limit is given by
\begin{eqnarray}
H_g= 2.72(g-z)_0 + 16.09.
\end{eqnarray}

The diagram in Fig. \ref{ppm} is very enlightening when we look at the density of objects. It is evident that there are two different populations within the sdAs: one to the red limit of the diagram and another in an intermediary region. While the distribution of the red population has no intersection with the known ELMs, the distribution resulting from the blue population shares colour properties with the known ELMs. Most of the ELMs in the blue distribution show $T\eff > 8\,000$~K (comprising sequences 1, 2, and 3 from Fig. \ref{fit_arrows}), while the red distribution contains objects mainly cooler than that (sequence 4 in Fig. \ref{fit_arrows}), explaining the two regimes which could be glimpsed in Fig. \ref{fit_arrows}. These distributions will be used to study the nature of these objects in terms of probabilities in Section \ref{probs}. We believe the red distribution is dominated by main sequence late-type stars, which can be found in the halo, with some possible contamination of cooler \mbox{(pre-)ELMs}, since there is an intersection with the blue distribution. The blue distribution, on the other hand, should contain the missing cool \mbox{(pre-)ELM} population, which is under-represented in the literature. Evolutionary models predict that ELMs spend about the same amount of time above and below $T\eff = 8\,500$~K, although the occurrence or not of shell flashes, which is still under discussion, can alter the time scales by even a factor of two. Either way, 20--50 per cent of the ELMs should show $T\eff < 8\,500$~K \citep{pelisoli2017}; however, as a systematic effect of the search criterion, only 4 per cent of the published ELMs are in this range.

\begin{figure}
	\includegraphics[angle=-90,width=\columnwidth]{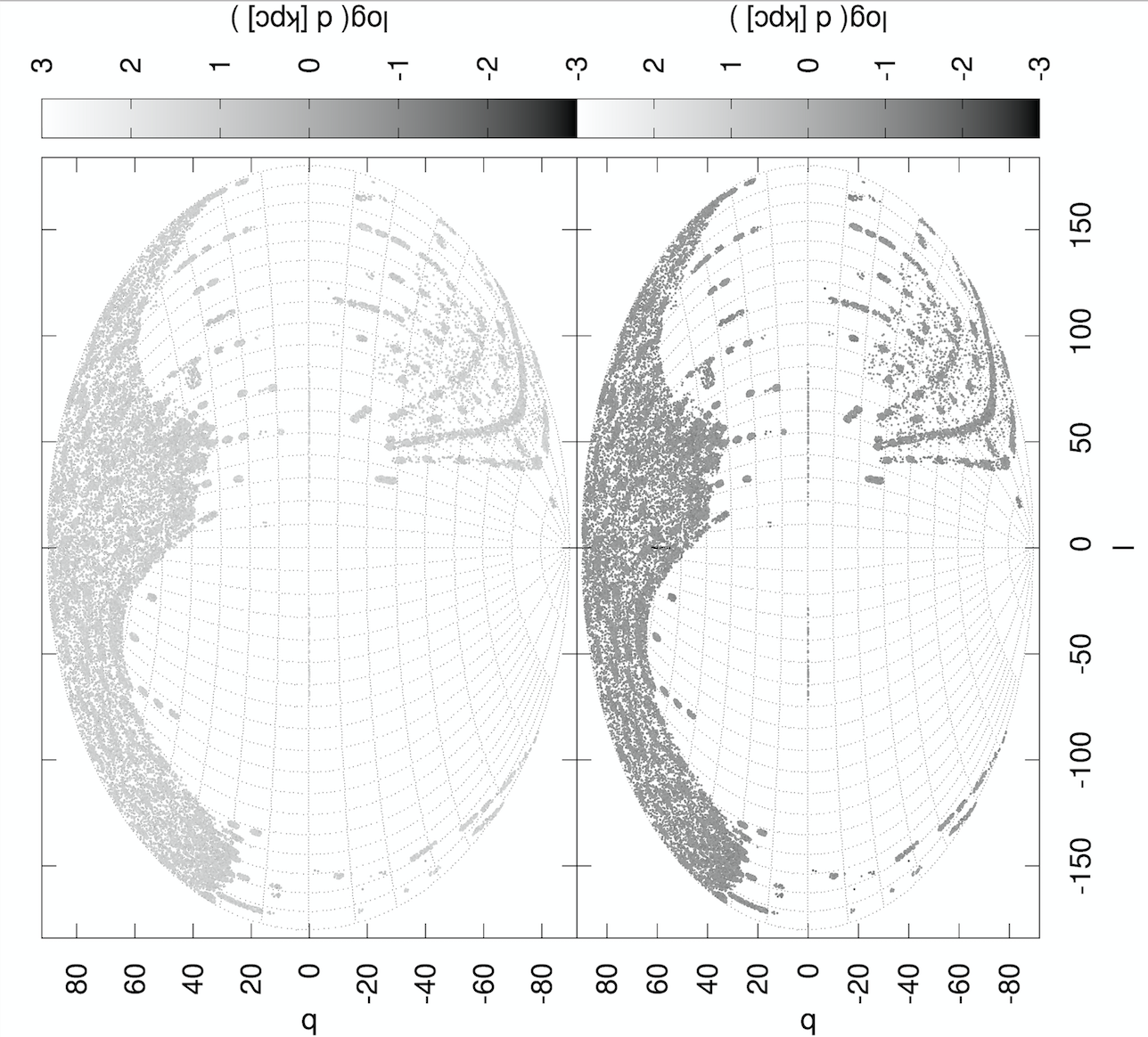}
    \caption{Aitoff projection showing the galactic latitude ($b$) and longitude ($l$) for all analysed objects. The grey scale indicates the estimated distance assuming either a main sequence radius (top panel) or a (pre-)ELM (bottom panel). There are no apparent streams, with the objects appearing to be distributed all over the SDSS footprint.}
    \label{aitoff}
\end{figure}

One of the outputs of our photometric fit, obtained by comparing the observed flux with the intensity given by the model, is the observed solid angle, related to the ratio between the radius $R$ of the object and its distance $d$. So assuming a radius, we can estimate the distance for the objects in our sample. We estimated the radii assuming both a main sequence radius, interpolated from a table with solar metallicity values given the estimated $T\eff$, and a (pre-)ELM radius interpolating evolutionary models. Fig. \ref{aitoff} shows an Aitoff projection of the position of the objects in sample A with colour-coded distance for both cases. About 2\,000 objects have estimated distances larger than 20~kpc when a main sequence radius is assumed; if indeed main sequence stars, it is unlikely they were formed within the Galactic disk, since there would not be enough time for them to migrate there within their evolutionary time. They could be accreted stars from neighbouring dwarf galaxies, but we have found no evidence of streams to support that. An alternative is that they are (pre-)ELMs white dwarfs. Most objects are contained within 3.0~kpc in this scenario.

\begin{figure}
	\includegraphics[angle=-90,width=\columnwidth]{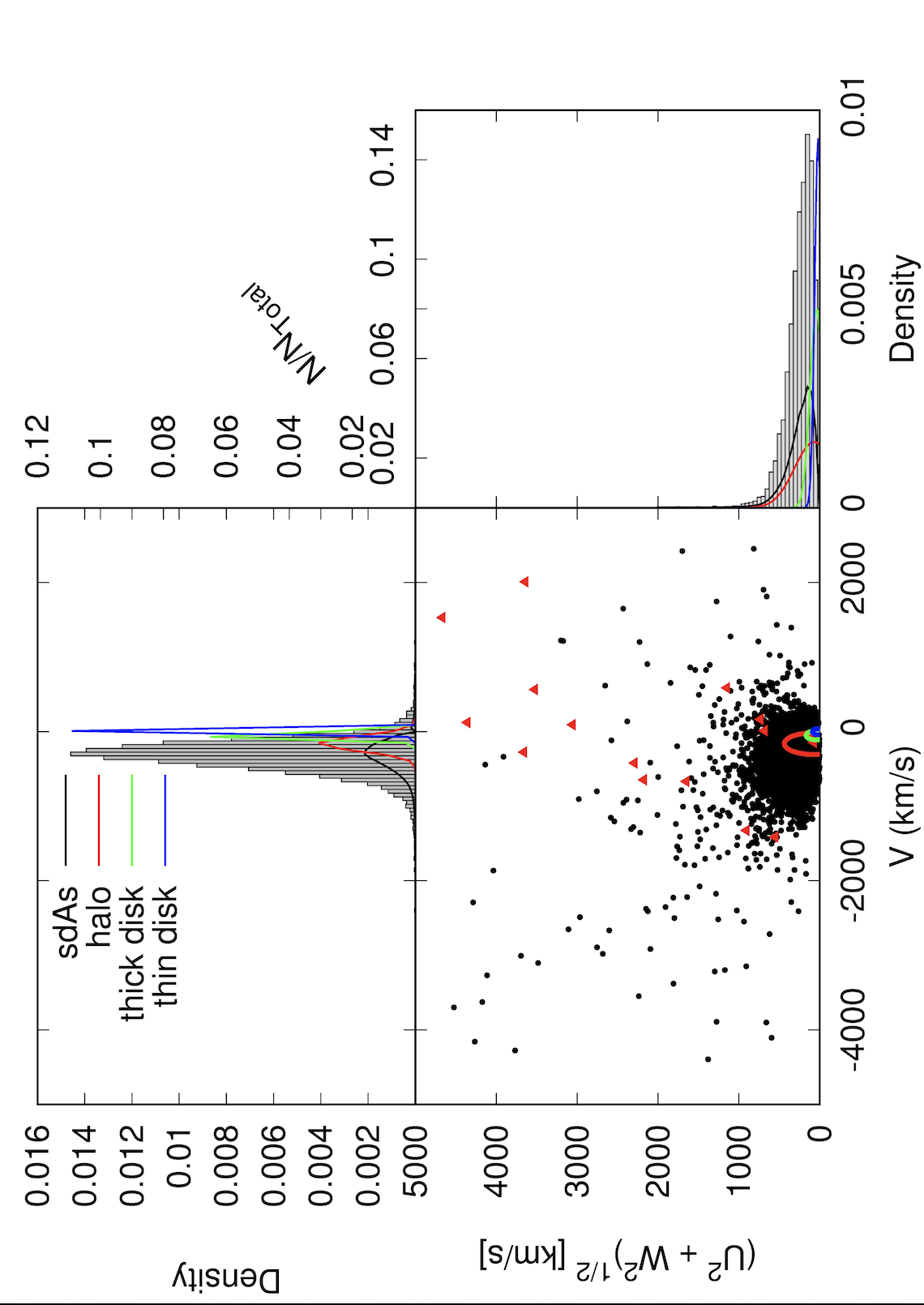}
    \caption{Toomre diagram of the objects in sample B, assuming a main sequence radius. Density plots are shown to left and on top. The ellipses indicate the $3\sigma$ values for halo (red), thick disk (green) and thin disk (blue) according to \citet{kordopatis2011}.}
    \label{uvw_MS}
\end{figure}

\begin{figure}
	\includegraphics[angle=-90,width=\columnwidth]{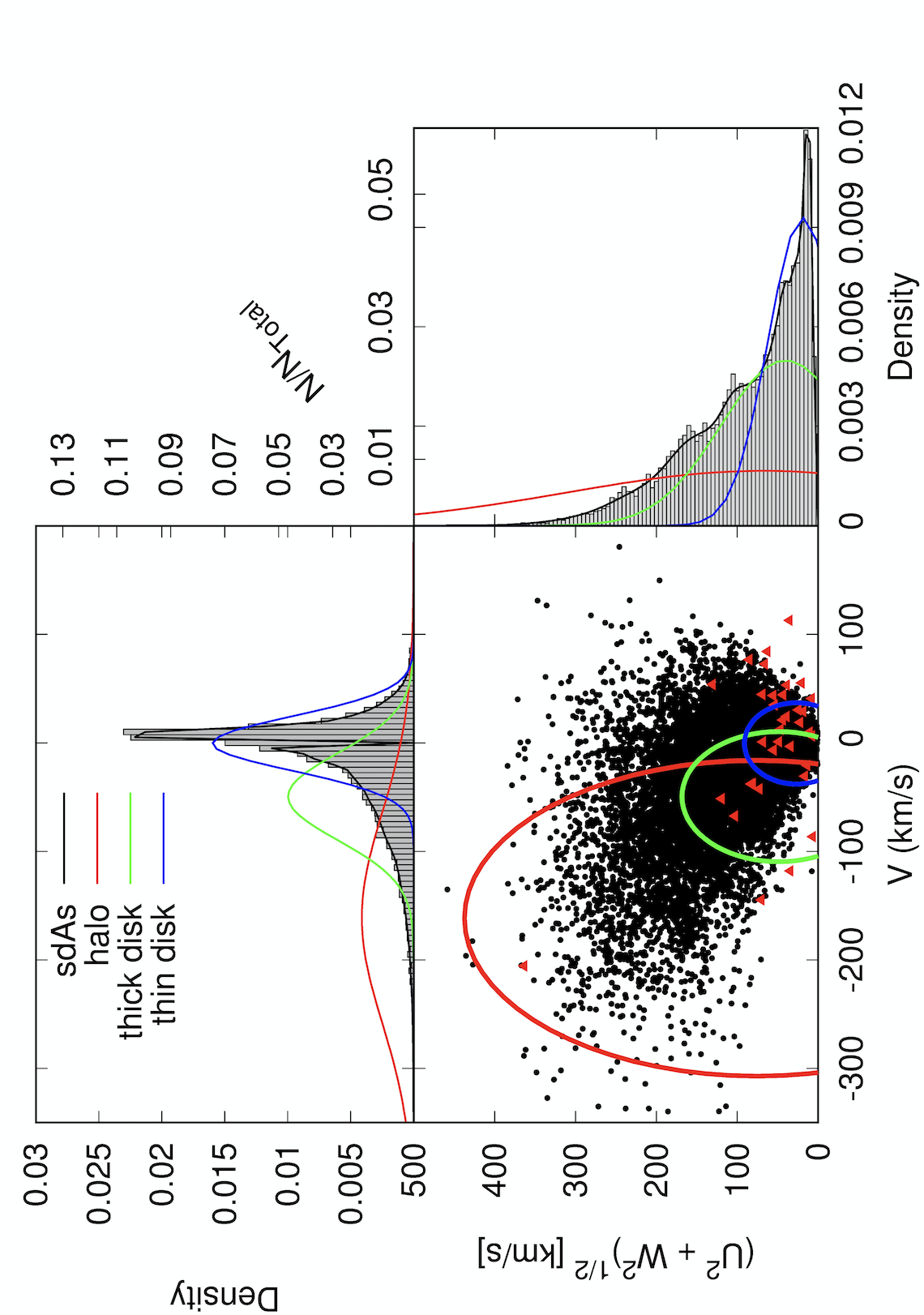}
    \caption{Same as Fig. \ref{uvw_MS}, but assuming a (pre-)ELM radius.}
    \label{uvw_ELM}
\end{figure}

Combining these distances with the reliable proper motions, and with radial velocities estimated from our spectral fit, we estimated the galactic velocities $U$, $V$, and $W$ \citep{johnson1987} given the main sequence or the ELM radius for sample B. The results are shown in Fig. \ref{uvw_MS} for a main sequence radius and in Fig. \ref{uvw_ELM} for a (pre-)ELM radius. About 38 per cent of the objects show velocities more than 3~$\sigma$ above the halo mean velocity dispersion when a main sequence radius is assumed --- implying a 1 per cent chance that they actually belong to the halo. Such high velocities also imply that the population cannot be related to blue horizontal-branch stars such as those studied by e.g. \citet{xue2008}. When the (pre-)ELM radius is assumed, on the other hand, the objects show a distribution consistent with a disk population. Further kinematic analysis, relying solely on the radial velocity component, can be found in Appendix \ref{furtherK}.

\section{Evolutionary Models}

\begin{figure*}
	\includegraphics[angle=-90,width=\textwidth]{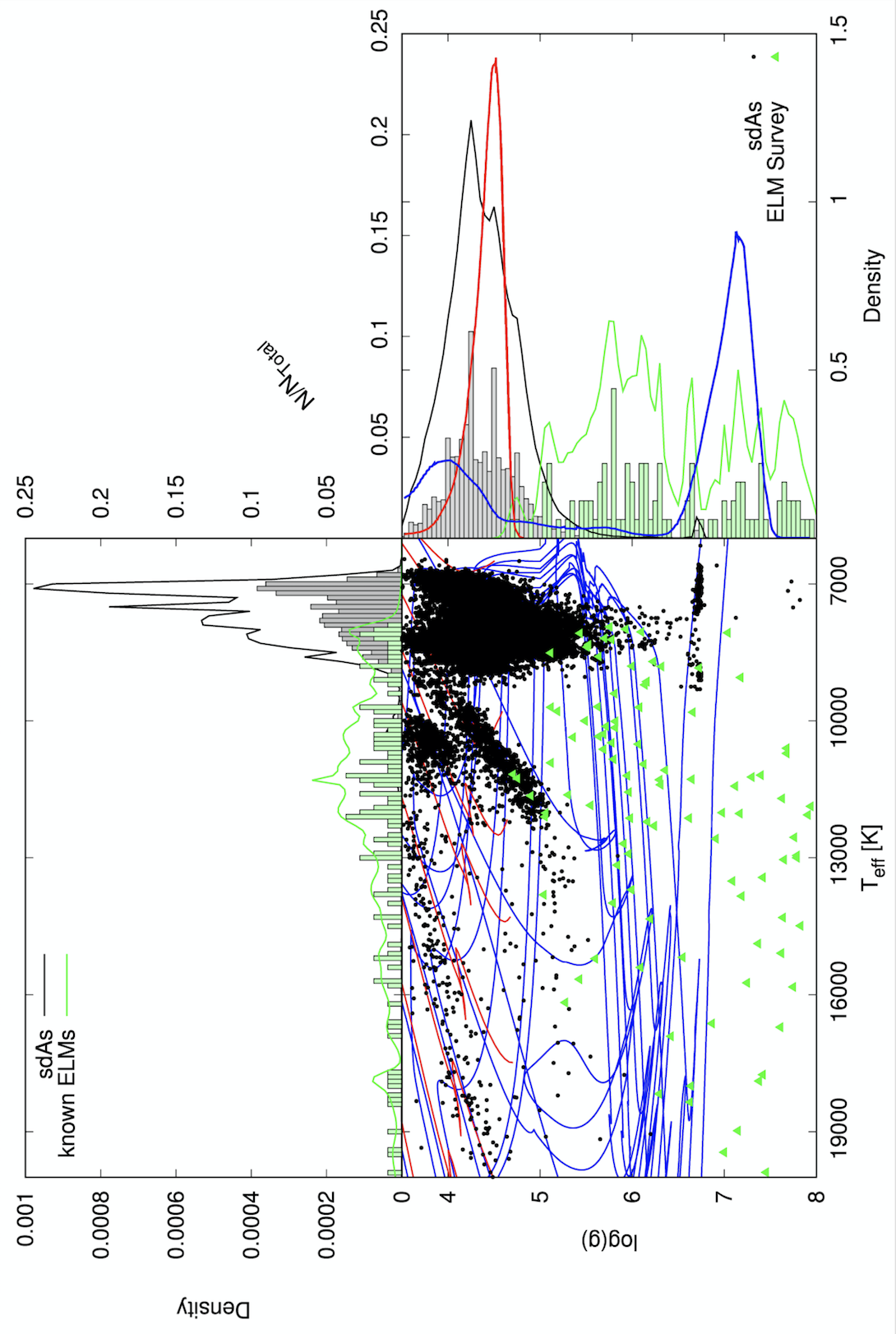}
    \caption{Bottom-left panel shows the $T\eff - \log g$ diagram for objects in sample A, shown as black dots, and known ELMs, shown as green triangles, compared to single evolution models of \citet{bertelli2008} and \citet{bertelli2009} (red), and binary evolution models of \citet{istrate2016} (blue). For $\log g \gtrsim 5.0$, the objects can only be explained by binary evolution tracks. The top panel shows the distributions in $T\eff$, while the bottom-right panel shows the distributions in $\log g$. The obtained probability distributions in $\log g$ for single (red) and binary evolution models (blue) are also shown. Note that there is significant overlap especially around $\log g \sim 4.5$. A colour version of this figure is available at the online version of the paper.}
    \label{evol}
\end{figure*}

We have also compared our obtained values of $T\eff$ and $\log g$ to predictions from evolutionary models, both to single and binary evolution, as shown in Fig. \ref{evol}. Single evolution models were taken from \citet{bertelli2008} and \citet{bertelli2009}. The plotted models are for $Z = 0.0001$, since the sdAs, if main sequence objects, should be in the halo, where the metallicity is low. We have used the binary evolution models of \citet{istrate2016}, the only ones to take rotation into account. The main panel in Fig. \ref{evol} shows that the values given by our solar metallicity fits are completely consistent with predictions from binary evolution models. Thus, given the values of $T\eff$ and $\log g$ alone, the sdAs could be explained as (pre-)ELMs.

Using both single and binary evolution models, we have also obtained two probability distributions for the $\log g$ given each evolutionary path. Our approach was to evaluate the time spent in each bin of $\log g$, $T_{\log g}$, for all single or binary models, over the total evolutionary time, $T_{\textrm{evol}}$, to obtain a probability $p_{\log g} = T_{\log g}/T_{\textrm{evol}}$ that an object resulting from single, or binary, evolution shows $\log g$ in the given bin.

To take into account the fact that brighter objects are more easily observable, even if short lived, this probability was combined with a volume correction assuming, by simplicity, spherical symmetry. Considering that a main sequence radius puts most of our objects in the halo, this approximation should hold. We have evaluated the observable volume $V_{\log g}$ of each bin of $\log g$ by summing up the volumes $V_{\textrm{obs}}$ of all models with $\log g$ in said bin, where the volume was calculated using the $M_v$ magnitude of each model, and assuming a saturation limit of $V = 14.5$ and a detection threshold of $V = 20.0$, so that
\begin{eqnarray}
\begin{split}
V_{\textrm{obs}} &=& \frac{4 \pi}{3} \left\{ \left [ 10^{(14.5-M_V+5)/5} \right ]^3 -\right. \\ & &\left.	 \left [ 10^{(20.0-M_V+5)/5} \right ]^3 \right\}
\end{split}
\end{eqnarray}
For each bin, we calculated the fraction of observable volume compared to the total volume throughout the evolutionary path, obtaining a probability of observing an object with the given $\log g$ during its evolution $p_{\textrm{obs}} = V_{\log g} / \sum V_{\log g}$. To obtain our final probability distribution, we have combined both probabilities in $1.0 - (1.0 - p_{\log g})\times(1.0 - p_{\textrm{obs}})$. The resulting distributions are shown in red for the main sequence models and in blue for the binary evolution models in Fig. \ref{evol}.

\section{A way-out: probabilities}
\label{probs}

Considering our previous analysis, it is clear that the observable properties of the sdAs are consistent with more than one evolutionary channel, since there is overlap between evolutionary paths. The only physical parameter that would allow an unique classification for the sdAs would be the radius, which, combined with $\log g$ estimates, would allow us to tell whether the objects have a degenerate nucleus. As there is no parallax measured for the sdAs, this will not be possible at least until Gaia's data release 2, scheduled for April 2018. High proper motion objects might not be in the data release 2, so the wait might be even longer. In the meantime, we can analyse the sdAs in terms of probability: do they have a higher probability of belonging to the main sequence, or can they be more easily explained by (pre-)ELMs? 

Based on the estimated $\log g$ compared to evolutionary models, on the reduced proper motion diagram distributions, and on the spacial velocities given either a (pre-)ELM radius or  a main sequence radius, we have estimated for each object in sample B a probability of belonging to the main sequence and a probability of being a \mbox{(pre-)ELM} star. Our intention is to provide a basis for future follow-up projects, such as time resolved spectroscopy, impossible with the present size of the samples, and to understand the sdA population as a whole.

The main sequence probability was evaluated taking into account three probabilities:
\begin{itemize}
\item[i)] probability of being explained by a single-evolution model (Fig. \ref{evol}) given the estimated solar abundance $\log g$: $p_{\textrm{\tiny MS1}}$;
\item[ii)] probability of belonging to the red distribution in Fig. \ref{ppm} given the $(g-z)_0$ colour : $p_{\textrm{\tiny MS2}}$;
\item[iii)] probability of belonging to the halo, thick or thin disk of the Galaxy given the U, V, W velocities estimated with a main sequence radius : $p_{\textrm{\tiny MS3}}$. 
\end{itemize}
The final probability was calculated as the complementary probability of the object \textit{not} belonging to the main sequence, assuming the intermediary probabilities listed above are independent:
\begin{eqnarray}
p_{\textrm{\tiny MS}} &=& 1 - (1 - p_{\textrm{\tiny MS1}}) \times (1 - p_{\textrm{\tiny MS2}}) \times (1 - p_{\textrm{\tiny MS3}}).
\end{eqnarray}

The (pre-)ELM probability on the other hand takes into account:
\begin{itemize}
\item[i)] probability of being explained by binary evolution models (Fig. \ref{evol}) given their estimated solar abundance $\log g$: $p_{\textrm{\tiny ELM1}}$;
\item[ii)] probability of belonging to the blue distribution in Fig. \ref{ppm} given the $(g-z)_0$ colour : $p_{\textrm{\tiny ELM2}}$;
\item[iii)] probability of belonging to the halo, thick or thin disk of the Galaxy given the U, V, W velocities estimated with a (pre-)ELM radius : $p_{\textrm{\tiny ELM3}}$. 
\end{itemize}
This gives a final probability, again assuming the intermediary probabilities are independent, of
\begin{eqnarray}
p_{\textrm{\tiny ELM}} &=& 1 - (1 - p_{\textrm{\tiny ELM1}}) \times (1 - p_{\textrm{\tiny ELM2}}) \times (1 - p_{\textrm{\tiny ELM3}}).
\end{eqnarray}

\begin{figure}
	\includegraphics[angle=-90,width=\columnwidth]{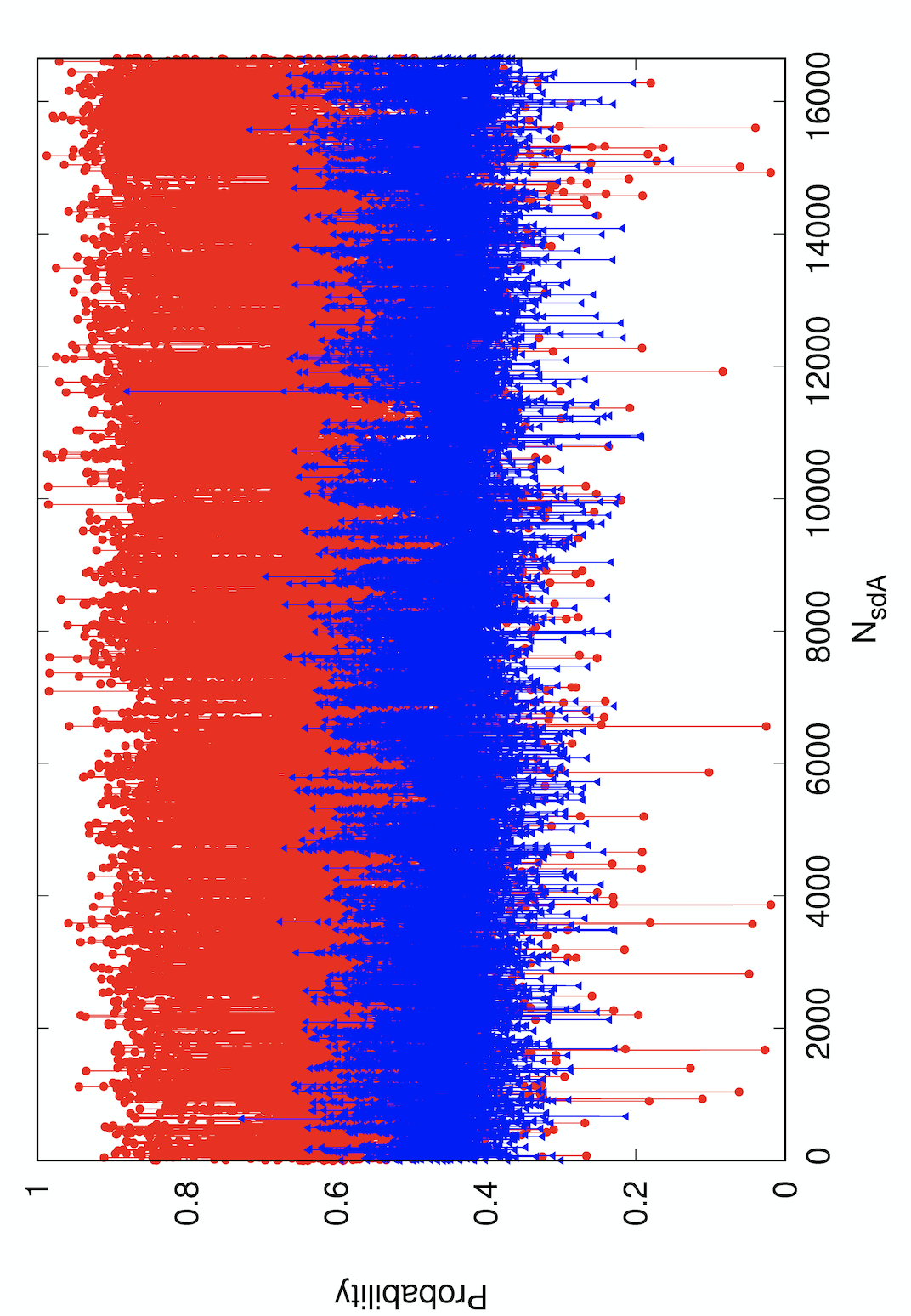}
    \caption{Circles in red show the probability of belonging to the main sequence, while triangles in blue show the resulting probability of being a (pre-)ELM object according to our evaluated distributions. The x-axis is simply a count of sdAs in sample B.}
    \label{probs1}
\end{figure}

\begin{figure}
	\includegraphics[angle=-90,width=\columnwidth]{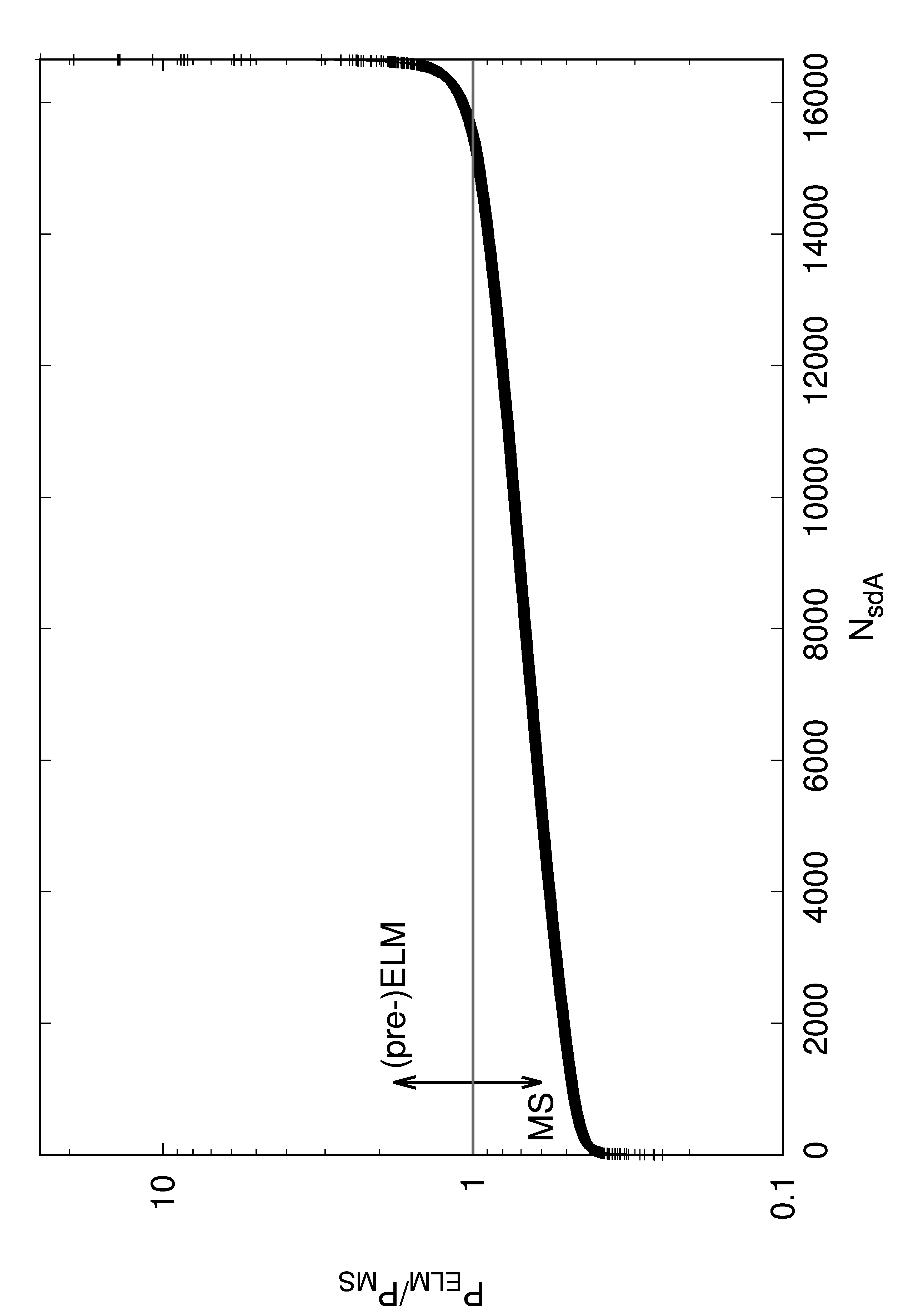}
    \caption{Probability of (pre-)ELM over probability of main sequence, ordered from smallest to largest. The y-axis is shown in log scale. Most objects do show a larger probability of belonging to the main sequence, but 1\,150 objects in sample B (about 7 per cent) are most probably (pre-)ELMs.}
    \label{probs2}
\end{figure}

As previously mentioned, there are intersections between the properties of the two populations, therefore the two probabilities are not independent and \textit{do not} sum up to one. Fig. \ref{probs1} shows the obtained results for the objects in sample B. If we analyse these results in terms of an average, a random sdA has 68 per cent probability of belonging to the main sequence, and a 46 per cent probability of being a (pre-)ELM. Doing the ratio between the two probabilities, we can obtain the objects which are more likely \mbox{(pre-)ELMs} in the sample. Fig. \ref{probs2} shows the (pre-)ELM probability over the main sequence probability. 1\,150 objects in sample B, or 7 per cent, have a higher probability of being (pre-)ELMs. They are listed in Table \ref{probs_table}. 170 of these objects show $\log g > 5.0$ --- implying they would be ELMs rather than pre-ELMs. Out of those, 146 also show $T\eff < 8\,500$~K.

Assuming all these objects have their nature correctly predicted, this would raise the number of ELMs with $T\eff > 8\,500$~K in the SDSS footprint to 97 (73 confirmed binaries of \citet{elmsurveyVII} in this range + 24 sdAs), while the number of objects with $T\eff < 8\,500$~K would be 149 (3 confirmed binaries of \citet{elmsurveyVII} in this range + 146 sdAs), making the cool ELM population about 50 per cent larger. The evolutionary models predict the same amount of time to be spent in both ranges, but shell flashes can reduce the hydrogen in the atmosphere, speeding up the cooling process and making it possible that the time spent at lower temperatures be higher by a factor of two, as found here. However, the circumstances where these shell flashes occur are still unclear. Follow-up of these objects to detect the true cool ELMs, allowing for an observational estimate of the rate of objects in the two $T\eff$ ranges, should be acquired to help calibrate the evolutionary models. We will discuss the current state of our follow-up in upcoming papers.

\begin{table*}
  \caption{Objects with a higher probability of being (pre-)ELM compared to main sequence. We show the obtained physical parameters, and the P-M-F of the spectrum from which they where derived, as well as the obtained probabilities for each case. The full table can be found in the on-line version of this paper.}
  \label{probs_table}
  \begin{tabular}{cccccc}
    \hline
    SDSS~J & P-M-F & $T\eff$ (K) & $\log g$ & $p_{MS}$ & $p_{ELM}$ \\
    \hline
    101053.89-004218.1 & 0270-51909-0161 &   7451(21) & 4.356(0.139) & 0.56 & 0.53 \\
    105752.38+001326.3 & 0276-51909-0599 &   6804(35) & 3.764(0.097) & 0.52 & 0.50 \\
    112941.67+000545.2 & 0281-51614-0117 &   7708(37) & 4.338(0.157) & 0.56 & 0.55 \\
    113704.83+011203.6 & 0282-51630-0561 &   8137(13) & 5.277(0.056) & 0.33 & 0.32 \\
    113704.83+011203.6 & 0282-51658-0565 &   8122(11) & 5.544(0.040) & 0.31 & 0.27 \\
    121213.83-003046.8 & 0287-52023-0240 &   7280(31) & 4.772(0.135) & 0.54 & 0.51 \\
    120811.97-004230.4 & 0287-52023-0311 &   8275(10) & 5.050(0.042) & 0.52 & 0.51 \\
    122000.93-005556.5 & 0288-52000-0100 &   7898(22) & 4.856(0.111) & 0.56 & 0.39 \\
    121715.08-000928.3 & 0288-52000-0234 &   8481(26) & 5.097(0.105) & 0.52 & 0.40 \\
    123222.47-001222.6 & 0289-51990-0028 &   7598(40) & 5.235(0.138) & 0.55 & 0.54 \\
    \hline
  \end{tabular}
\end{table*}

\section{Summary \& Conclusions}

We have analysed a sample of narrow-line hydrogen spectra identified in the SDSS, estimating $T\eff$ and $\log g$ from their spectra using new spectral models derived from solar abundance atmospheric models. Comparing these results to previous pure-hydrogen models by \citet{dr12cat}, we showed that the shift in $\log g$ when metals are added is not a constant, but depends on $\log g$, unlike what was suggested by \citet{brown2017}. For objects with a pure-hydrogen $\log g > 5.5$ though, as the objects analysed by \citet{brown2017}, the pure-hydrogen $\log g$ seems in fact to be about 1~dex higher than the solar abundance $\log g$.

With these new models, we have identified new sdAs in the range $5.0 \leq \log g \leq 7.0$ and $7\,000 \leq T\eff < 20\,000$~K. We analyse the colours of the whole sample of narrow-line hydrogen spectra, obtaining that the spectroscopic $\log g$ does not seem to agree with the position of the objects in colour-colour diagrams. This might indicate that there is still missing physics in the models; the addition of metals alone does not solve the discrepancy. Other missing opacities, such as molecular contributions, might be the explanation. However, the discrepancy could also be solved if the reddening is underestimated. Although out of the scope of this work, we consider that both possibilities should be investigated. One key-result obtained from the colour analysis is that the sdAs can not be explained as binaries of a hot subdwarf with a main sequence star, since they do not show significant flux in the UV. There is also no indication of infrared excess for over 98 per cent of the sample.

The most significant result is, though, that the sdAs are clearly composed of two populations. One population contains the red objects, and it has no overlap with the known ELMs. These could be explained as cool late-type main sequence stars, when the velocities are consistent with the halo population. On the other extreme, there is a blue population, which does overlap with known ELMs, but contains cooler objects. Considering that there is still a missing cool ELM population to be found, given the predictions of evolutionary models, it is very likely that these objects belong to the blue population of sdAs.

Analysing the estimated distances and spacial velocities for the objects, we obtain that over 35 per cent of them show too high velocities to belong to the halo when a main sequence radius is assumed. These objects cannot therefore be explained as simply metal-poor main sequence stars of type A--F. The discrepant velocities are solved when a \mbox{(pre-)ELM} radius is assumed for these objects, in which case their velocities become consistent with the disk distribution. Some percentage of these objects might be of binary stars, such as blue stragglers, in which case the velocities could be explained as orbital instead of spacial motion. A better sense of the nature of this population will be obtained when their parallax is released by Gaia. What we should keep in mind is that, given their apparent extreme velocities and distances, they certainly can help us study the dynamics of the halo.

We have also compared our estimated values of $T\eff$ and $\log g$ to evolutionary models, both single and binary. A very interesting result is that the parameters for the objects in our sample are consistent with those expected from binary evolution models. Considering the time spent in each bin of $\log g$ and the brightness at such phases, even pre-ELMs with $\log g<4.0$ have considerable probability of being observed.

Taking into account the derived probabilities from the evolutionary models, combined with the probabilities given the colours and spacial velocities, we have estimated probabilities for each object to be either a main sequence star or a (pre-)ELM. As there are significant overlap between the parameters of each class, the probabilities do not sum up to one. Comparing the probabilities, we find that about 7 per cent of the sdAs are better explained as (pre-)ELMs than as main sequence stars, a much larger percentage than found by \citet{brown2017} studying a small sample of eclipsing stars. Considering the physical parameters of the objects with a higher probability of being (pre-)ELMs, our result is consistent with the existence of two times as many cool ELMs ($T\eff < 8\,500$~K) as hot ELMs. However, as in many cases the probability of being ELM is only marginally larger than of belonging to the main sequence, this result should be confirmed by follow-up of these objects, as we will discuss in upcoming papers.

Even if only a small percentage of the sdAs is composed by (pre-)ELMs, the number is high enough to potentially double the number of known ELMs. The cool ELM population, in particular, seems to be within the sdAs. As our models for this kind of object are still under development, monitoring of the sdAs is essential to find this missing population. Tables \ref{Tlog_table} and \ref{probs_table} published here are a valuable asset to guide this monitoring. Our understanding of binary evolution, and especially of the common envelope phase that ELMs must experience, can be much improved if we have a sample covering all parameters predicted by these models. The sdA sample provides that.

Our understanding of the formation and evolution of the Galactic halo would also benefit from more detailed study of the sdAs. Many seem to be in the halo with ages and velocities not consistent with the halo population. It is possible that accreted stars from neighbouring dwarf galaxies might be among them. Those whose velocities are in fact consistent with the halo can in turn help us study its dynamics and possibly better constrain the gravitational potential of the halo. Comparing the numbers of both populations, we can obtain clues of how different formation scenarios, namely accretion and formation in locus, contributed to the halo.

The key message of our results is that we should not overlook the complexity of the sdAs. They are of course not all pre-ELM or ELM stars, but they cannot be explained simply as main sequence metal-poor A--F stars. They are most likely products of binary evolution and as such are a valuable asset for improving our models.

\section*{Acknowledgements}

IP and SOK acknowledge support from CNPq-Brazil. IP was also supported by Capes-Brazil under grant 88881.134990/2016-01. DK received support from programme Science without Borders, MCIT/MEC-Brazil. We thank Thomas G. Wilson for useful discussions on IR-excess, Alejandra D. Romero for providing ZAHB tables, and the anonymous referee for their helpful comments and suggestions.

This research has made extensive use of NASA's Astrophysics Data System.

Funding for the Sloan Digital Sky Survey IV has been provided by the Alfred P. Sloan Foundation, the U.S. Department of Energy Office of Science, and the Participating Institutions. SDSS-IV acknowledges
support and resources from the Center for High-Performance Computing at
the University of Utah. The SDSS web site is www.sdss.org.

SDSS-IV is managed by the Astrophysical Research Consortium for the 
Participating Institutions of the SDSS Collaboration including the 
Brazilian Participation Group, the Carnegie Institution for Science, 
Carnegie Mellon University, the Chilean Participation Group, the French Participation Group, Harvard-Smithsonian Center for Astrophysics, 
Instituto de Astrof\'isica de Canarias, The Johns Hopkins University, 
Kavli Institute for the Physics and Mathematics of the Universe (IPMU) / 
University of Tokyo, Lawrence Berkeley National Laboratory, 
Leibniz Institut f\"ur Astrophysik Potsdam (AIP),  
Max-Planck-Institut f\"ur Astronomie (MPIA Heidelberg), 
Max-Planck-Institut f\"ur Astrophysik (MPA Garching), 
Max-Planck-Institut f\"ur Extraterrestrische Physik (MPE), 
National Astronomical Observatories of China, New Mexico State University, 
New York University, University of Notre Dame, 
Observat\'ario Nacional / MCTI, The Ohio State University, 
Pennsylvania State University, Shanghai Astronomical Observatory, 
United Kingdom Participation Group,
Universidad Nacional Aut\'onoma de M\'exico, University of Arizona, 
University of Colorado Boulder, University of Oxford, University of Portsmouth, 
University of Utah, University of Virginia, University of Washington, University of Wisconsin, 
Vanderbilt University, and Yale University.

%%%%%%%%%%%%%%%%%%%%%%%%%%%%%%%%%%%%%%%%%%%%%%%%%%



\bibliographystyle{mnras}
\bibliography{sdAs_paper1} % if your bibtex file is called example.bib

%\begin{thebibliography}{99}
%\end{thebibliography}

%%%%%%%%%%%%%%%%%%%%%%%%%%%%%%%%%%%%%%%%%%%%%%%%%%

%%%%%%%%%%%%%%%%% APPENDICES %%%%%%%%%%%%%%%%%%%%%

\appendix

\section{Comparison with parameters from the SDSS pipelines}
\label{more_models}

\begin{figure*}
	\includegraphics[width=\textwidth]{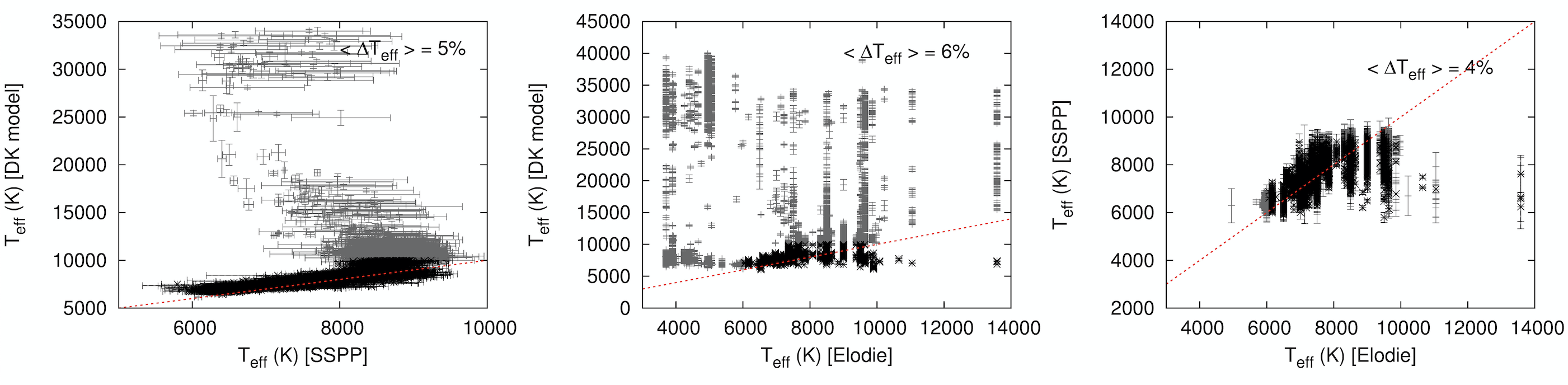}
    \caption{Comparison between our estimate of $T\eff$ (DK model) and both pipelines, as well as the comparison between the pipelines. Grey dots are outside of the area covered by SDSS grids, while black dots are within it. The red dashed line represents equality. The average difference for estimates within both grids is shown in each plot. It is clear that there is no discrepancy in the region covered by both grids. Our higher temperatures are backed up by both the GALEX ultraviolet flux, and the SDSS classification as type O and B, which is not coherent with their own estimated temperature.}
    \label{comp_teff}
\end{figure*}

\begin{figure*}
	\includegraphics[width=\textwidth]{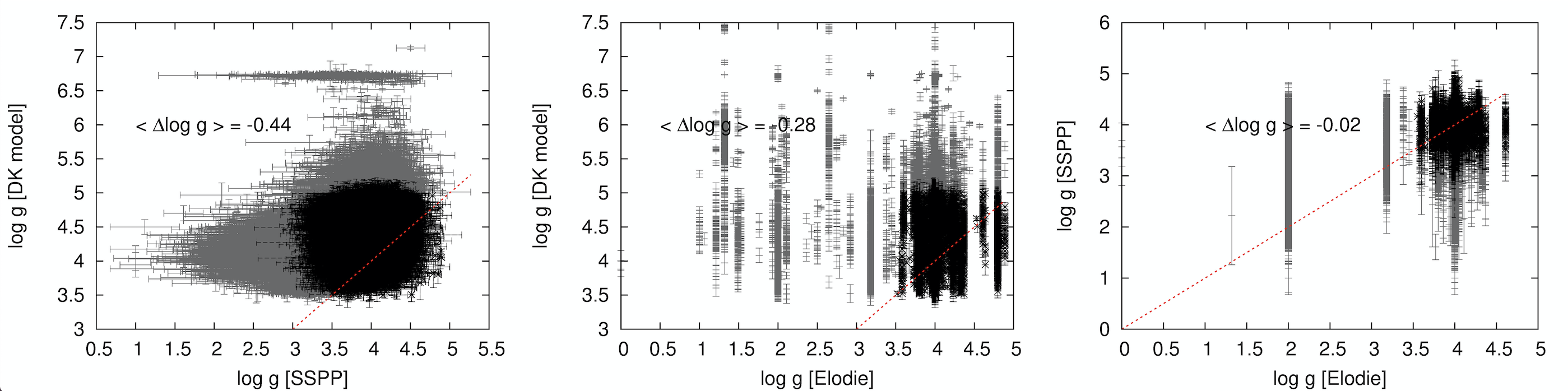}
    \caption{Comparison between our $\log g$ and that of the pipelines. Colours are the same as in Fig. \ref{comp_teff}. The spread is larger and our fit favours, in average, slightly higher $\log g$ values, given the extension of our grid to higher values.}
    \label{comp_logg}
\end{figure*}

The comparison between our estimated parameters and those given by the SDSS pipelines is not straightforward, as the model grids do not cover the same range of $T\eff$ and $\log g$. We include this analysis here as an appendix to illustrate their differences. We do not, however, consider such comparison a valid test of our models, since the SDSS pipeline grids are strongly incomplete in terms of $T\eff$ and $\log g$. Figs. \ref{comp_teff} and \ref{comp_logg} show the comparison between the $T\eff$ and the $\log g$, respectively, that we obtained for the objects with a good fit (sample A) and the values given by two SDSS pipelines, the SEGUE Stellar Parameter Pipeline (SSPP) \citep{lee2008} and the best match from the ELODIE stellar library \citep{prugniel2001}. The two pipelines are also compared. We find good agreement ($\sim$ 5 per cent) in $T\eff$ between our fit and both pipelines. Our $\log g$ is $0.44$ higher than the SSPP estimate, and $0.28$ higher than the Elodie values, which is probably due to the difference in the extension of the grids. Moreover, this average shift is not enough to raise the $\log g$ of a typical main sequence A star ($\log g \sim 4.3$) to values above 5.0.

\begin{figure}
	\includegraphics[angle=-90,width=\columnwidth]{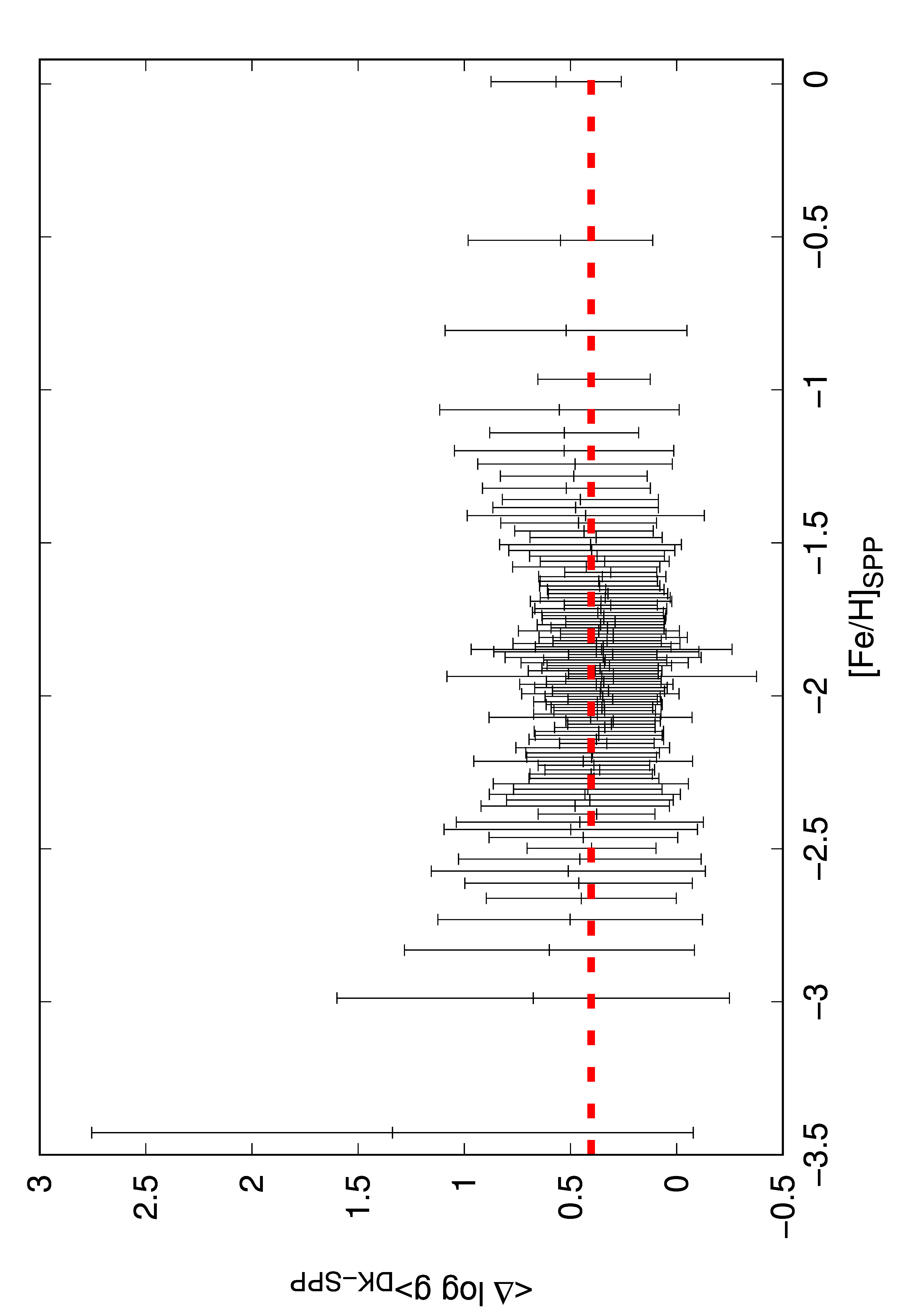}
	\caption{Absolute difference in $\log g$ between our models and the SSPP determination, averaged every 100 points, as a function of the metallicity $[Fe/H]$ given by SSPP. The red dashed line shows the overall average. There seems to be no strong dependence of the difference between surface gravity on the metallicity.}
	\label{metallicity}
\end{figure}

We have also performed an analysis to estimate whether the metallicity would play a significant role in the gravity estimate. In order to do that, we verified if the difference in $\log g$ between our determination and that of SSPP was dependent on the value of $[Fe/H]$ given by SSPP. There are 10120 objects in sample A with SSPP determinations. We ordered them by $[Fe/H]$, and calculated the average of the absolute difference in $\log g$, $\sqrt{ (\log g_{\textrm{DK}} - \log g_\textrm{SSPP} )^2}$, as well as the standard deviation of this average, every 100 points. Fig. \ref{metallicity} suggests that the metallicity is not a dominant uncertainty factor in the gravity estimate, since there is no dependence of the difference between determinations on $[Fe/H]$.

\section{Further Kinematic Analyses}
\label{furtherK}

\begin{figure}
	\includegraphics[angle=-90,width=\columnwidth]{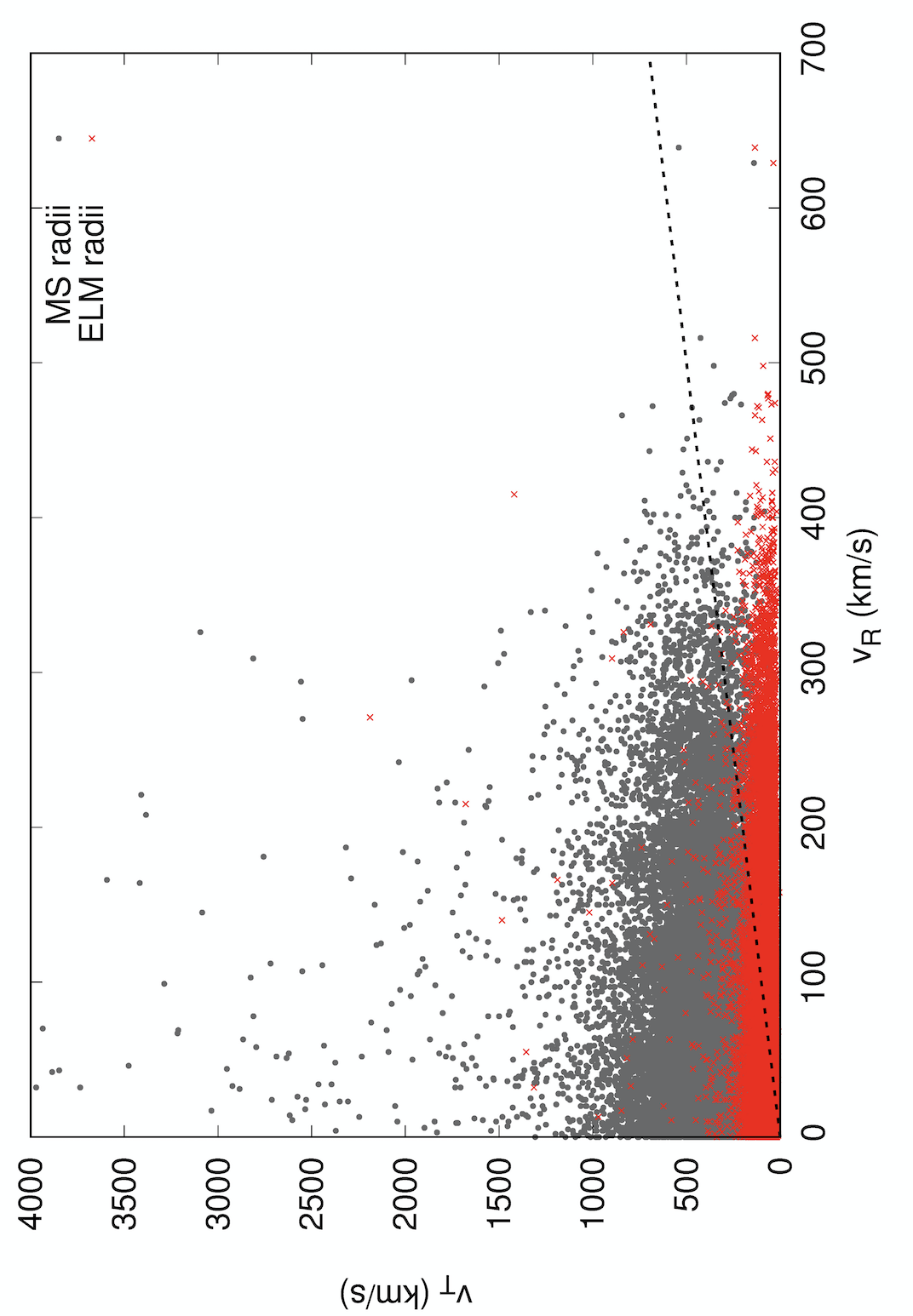}
	\caption{Comparison between the radial velocity estimated from SDSS spectra ($v_R$), and the transversal velocity derived from the GPS1 proper motion given the estimated distance ($v_T$), both assuming a main sequence (grey dots) and an ELM radius (red crosses) for sample B only. The spacial motion is dominated by the tangential velocity if a main sequence radius is assumed, with many objects showing $v_T$ much above the escape velocity. That does not happen when an ELM radius is assumed, with very few exceptions.}
	\label{vRvT}
\end{figure}

\begin{figure}
	\includegraphics[width=\columnwidth]{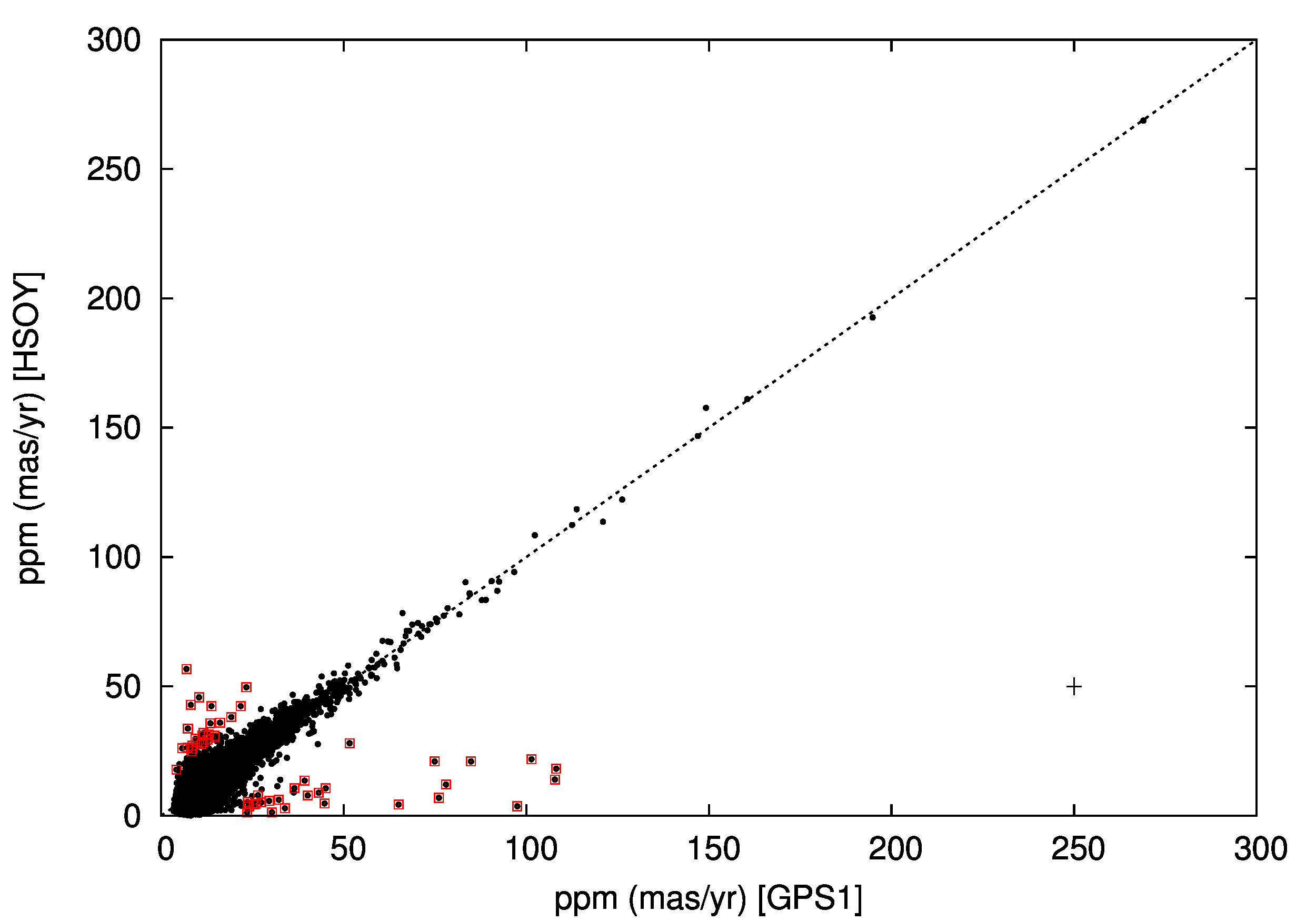}
	\caption{Comparison between the GPS1 and HSOY proper motions. The outliers are marked with red squares. The dashed line represents equality between the two values. Average uncertainties are shown as a cross in the bottom right of the panel.}
	\label{crossmatch}
\end{figure}

In many cases, the space motions of objects in sample B are dominated by the transversal velocity component, especially when a main sequence radius is assumed, as shown in Fig. \ref{vRvT}. To verify this was not a consequence of outliers in the proper motion catalogue \citep[such as found by e.g.][]{ziegerer2015}, we cross-matched the GPS1 proper motions with both the Hot Stuff for One Year catalogue \citep[HSOY,][]{altmann2017}, and the proper motions given in the SDSS tables \citep{munn2004,munn2014}. Only 69 objects from sample B differ by more than 3-$\sigma$ when comparing HSOY and GPS1 (see Fig. \ref{crossmatch}); 110 when we compare GPS1 to Munn et al. Such numbers are not of concern given the sample size, hence the objects were kept in the sample.

\begin{figure}
	\includegraphics[angle=-90,width=\columnwidth]{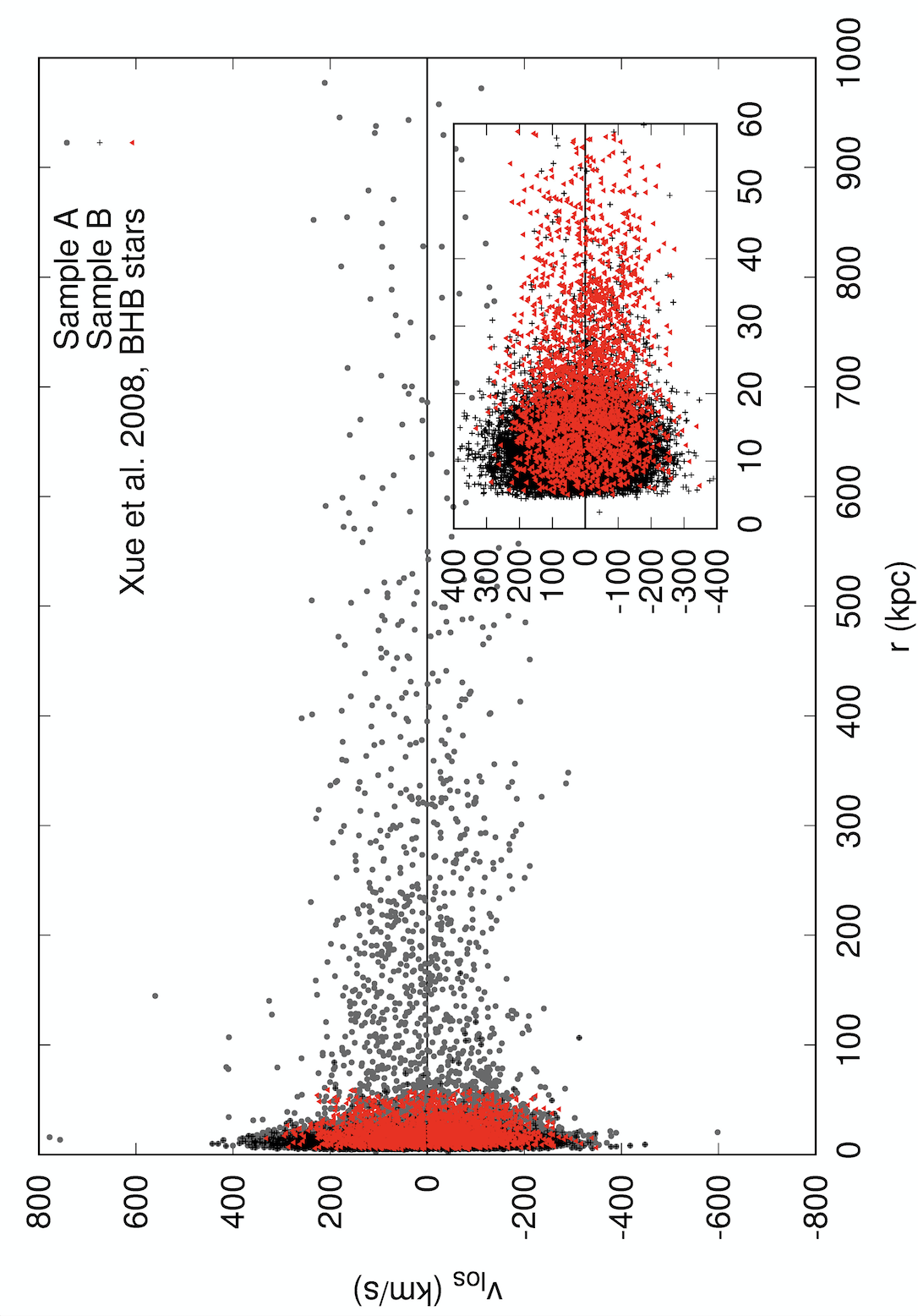}
	\caption{Line-of-sight velocity in the GSR as a function of Galactocentric distance. Sample A (grey dots) and B (black crosses) are of course different, with the unreliable proper motion of sample A giving in some cases unrealistic values of distance. Focusing on sample B only (smaller overlay panel), distances and velocities are on a similar range as halo BHB stars (shown as red triangles for comparison), with slightly larger velocities giving a somewhat higher dispersion, as can be seen in Fig. \ref{vlos_hist}.}
	\label{vlos_r}
\end{figure}

\begin{figure}
	\includegraphics[angle=-90,width=\columnwidth]{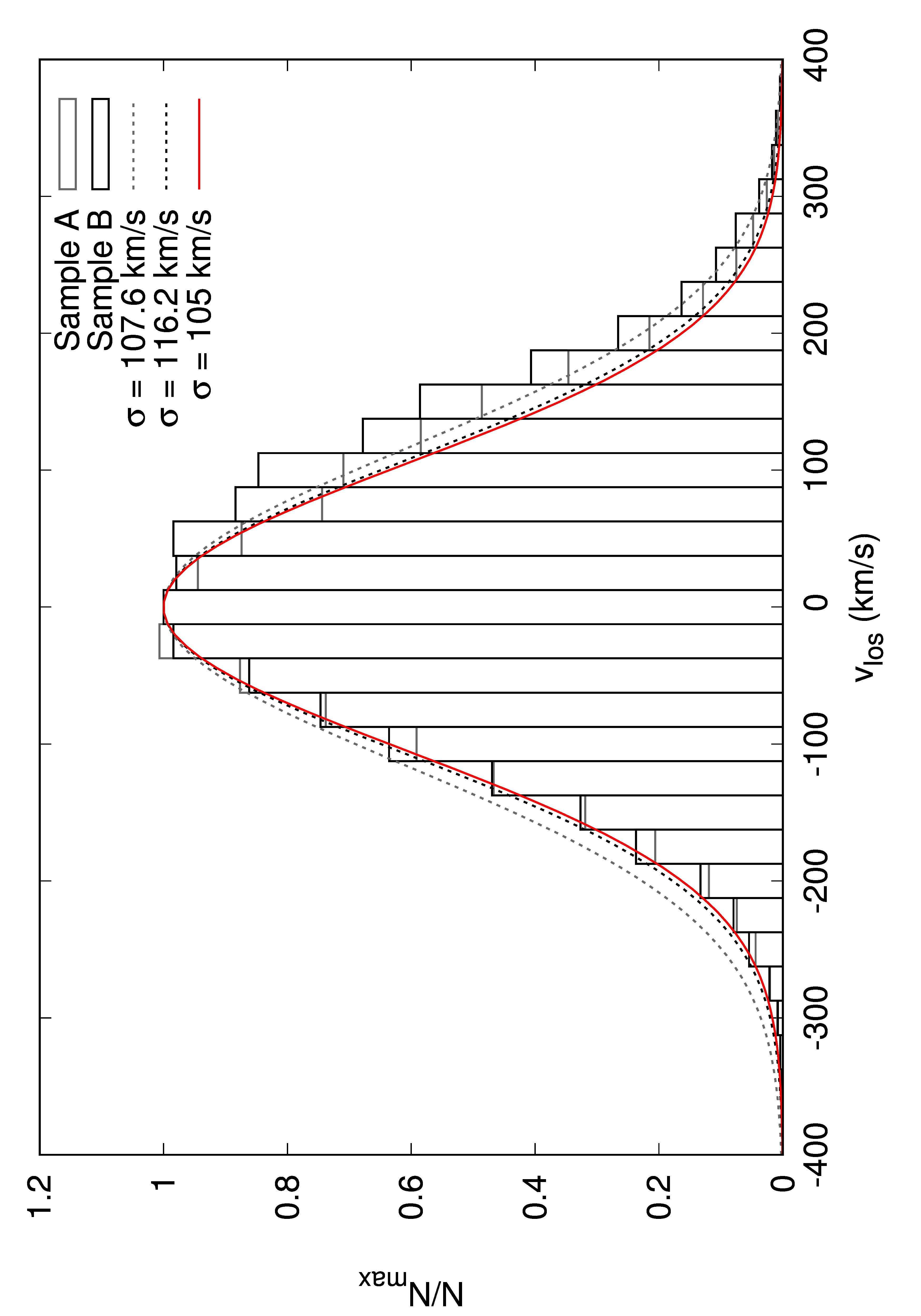}
	\caption{Distribution of $v_{los}$ for both samples A (black) and B (grey). The best fit Gaussian for each sample is shown as a dashed line of same colour. A Gaussian of width 105~km/s, similar to the value found by \citet{xue2008} for halo BHB stars is shown in red for comparison. Samples A and B show very similar dispersions, which are consistent with the estimated for the Galactic halo.}
	\label{vlos_hist}
\end{figure}

To work around possible effects risen by inaccuracy in the transversal velocity component, we have also performed a kinematic study relying on the radial velocity estimate alone. We have computed the Galactocentric distance ($r$) and the of line-of-sight velocities ($v_{los}$) in the Galactic standard of rest (GSR) frame following equations 4 and 5 of \citet{xue2008}. For the Galactocentric distance, we have assumed a MS radius. Fig. \ref{vlos_r} shows $v_{los}$ as a function of $r$, for both samples A and B, compared to the BHB stars of \citet{xue2008}. The unreliable proper motions of sample A result on unrealistic distance estimates, which are avoided by focusing on sample B, which shows similar distances to the halo BHBs. It is important to emphasise, however, that most sdAs show lower temperature than the ZAHB at their $\log g$, as shown in Fig. \ref{comp}. In Fig. \ref{vlos_hist}, we show the distribution of $v_{los}$ for both sample A and sample B, compared to a Gaussian of width $\sigma = 105$~km/s, as found by \citet{xue2008} for BHB halo stars. There is no significant difference between the distributions of samples A and B. Moreover, both show a dispersion of the same order of the halo stars studied by \citet{xue2008} when a main sequence radius is assumed, hence the conclusion that, if indeed main sequence stars, the sdAs would have to be in the halo is not dependent on the tangential velocity estimate.

\section{List of Possible Hot Subdwarfs}

\begin{table}
  \caption{Objects for which our fit indicated $T\eff > 20\,000$~K and $\log g < 6.5$, which are possible hot subdwarfs not listed in \citet{geier2017}.}
  \label{sdBs}
  \begin{tabular}{cc}
    \hline
    SDSS~J & P-M-F \\
    \hline
    000138.22+282731.9 & 2824-54452-0443 \\
	003739.05+000842.3 & 3588-55184-0616 \\
	013940.56+003626.6 & 1907-53265-0339,1907-53315-0335 \\
	021419.64-084505.3 & 4395-55828-0478 \\
	062336.21+642730.0 & 2301-53712-0627 \\
	063659.39+832025.0 & 2548-54152-0076 \\
	064846.81+373614.3 & 2700-54417-0079 \\
	073225.82+153729.7 & 2713-54400-0172 \\
	073546.72+410350.5 & 2701-54154-0541 \\
	073654.60+280923.4 & 4456-55537-0728 \\
	075029.26+181749.5 & 2729-54419-0428 \\
	075640.08+071806.6 & 4843-55860-0525 \\
	080520.22+000944.5 & 4745-55892-0147 \\
	081544.25+230904.7 & 4469-55863-0004 \\
	082606.25+113913.8 & 4508-55600-0855 \\
	083830.10+135117.6 & 4500-55543-0263 \\
	090141.48+345924.4 & 4645-55623-0456 \\
	091301.01+305119.8 & 2401-53768-0389 \\
	091721.87+283656.0 & 5797-56273-0546 \\
	091914.64+480306.0 & 5813-56363-0709 \\
	093946.04+065209.4 & 1234-52724-0304 \\
	100233.49+164500.5 & 5326-56002-0106 \\
	100442.16+132711.6 & 5328-55982-0828 \\
	104159.64+192254.5 & 5886-56034-0628 \\
	104437.73+145213.3 & 5350-56009-0768 \\
	105847.65+203917.4 & 5876-56042-0729 \\
	111917.41+050617.3 & 4769-55931-0682 \\
	112121.98+453955.5 & 6648-56383-0750 \\
	113312.12+010824.8 & 2877-54523-0369 \\
	121123.36+611203.8 & 0954-52405-0567,6972-56426-0278 \\
	121910.44+230020.7 & 5979-56329-0474 \\
	132138.31+133200.1 & 5427-56001-0746 \\
	132210.94+421216.9 & 6622-56365-0473 \\
	133540.82+014725.2 & 4045-55622-0014 \\
	134531.00-005314.4 & 4043-55630-0180 \\
	140603.27+374216.6 & 4711-55737-0208 \\
	141055.68+374340.6 & 4712-55738-0466 \\
	141810.60-020038.8 & 4032-55333-0532,4035-55383-0990 \\
	150421.06-005613.8 & 4020-55332-0950 \\
	151404.97+065352.7 & 2927-54621-0268 \\
	151940.51+014457.1 & 4011-55635-0310 \\
	152729.40+222448.5 & 3954-55680-0244 \\
	153049.60+321425.4 & 4723-56033-0639 \\
	155105.64+452134.0 & 3454-55003-0204 \\
	155241.28+045428.7 & 4877-55707-0254 \\
	160612.98+521919.3 & 2187-54270-0224 \\
	161430.90+041843.4 & 2189-54624-0633 \\
	164014.57+320325.2 & 5202-55824-0900 \\
	165237.92+240302.8 & 4181-55685-0236 \\
	165406.79+271117.7 & 4185-55469-0985 \\
	165634.74+231256.4 & 3290-54941-0378 \\
	170126.69+204620.5 & 4175-55680-0810 \\
	172037.66+534009.3 & 0359-51821-0273 \\
	191837.28+370917.5 & 2821-54393-0203 \\
	204403.97-051135.6 & 0635-52145-0360 \\
	204802.45+002753.1 & 1116-52932-0512 \\
	211716.97-005401.6 & 4192-55469-0376 \\
	213301.81-004914.7 & 4194-55450-0178 \\
	215014.24+233039.1 & 5953-56092-0185 \\
	224145.04+292426.0 & 6585-56479-0883 \\
	225654.02+074449.8 & 2325-54082-0376 \\
	232810.27-084156.3 & 3145-54801-0371 \\
    \hline
  \end{tabular}
\end{table}


%%%%%%%%%%%%%%%%%%%%%%%%%%%%%%%%%%%%%%%%%%%%%%%%%%

% Don't change these lines
\bsp	% typesetting comment
\label{lastpage}
\end{document}